\documentclass[letterpaper,twocolumn,10pt]{article}

\usepackage{usenix-2020-09}

\def\showcomments{0}

\usepackage{url}

\usepackage{graphicx}
\usepackage{diagbox}

\usepackage{xspace}
\usepackage{xcolor} 
\usepackage{tabularx}
\usepackage{subcaption}
\usepackage{enumitem}
\usepackage[T1]{fontenc}
\usepackage{hyperref}
\usepackage{xurl}
\hypersetup{breaklinks=true}
\usepackage{amsmath}
\usepackage{tikz}
\usepackage{soul}
\usepackage{multirow}
\usepackage{booktabs}
\usepackage{csquotes} 
\usepackage[normalem]{ulem} 
\usepackage[capitalize,noabbrev]{cleveref}

\usepackage{listings}
\usepackage{siunitx} 
\usepackage{pdflscape}
\usepackage{pgfplotstable}
\usepackage{pgfplots}
\usepackage{pgfplotstable}
\usepgfplotslibrary{colorbrewer}
\usetikzlibrary{patterns}

\newcommand{\bheading}[1]{{\vspace{4pt}\noindent\textbf{#1}}}

\ifnum\showcomments=1
    \newcommand{\gennote}[4][blue]{\textcolor{#1}{$\rule{8pt}{8pt}_{\textsf{\scshape\bfseries #2}}$ \textcolor{gray}{\emph{\sout{#3}}}#4}}
    \newcommand{\liangw}[1]{{\textcolor{red}{[LW: #1]}}}
    \newcommand{\henry}[1]{\gennote[green!40!black]{henry}{}{[#1]}}
    
    \newcommand{\joel}[1]{\gennote[orange]{joel}{}{[#1]}}
    \newcommand{\note}[1]{\gennote{note}{}{[#1]}}
    \newcommand{\todo}[1]{\gennote[red]{todo}{}{[#1]}}
    \newcommand{\adrian}[1]{\gennote[blue]{adrian}{}{[#1]}}
    \newcommand{\prateek}[1]{\gennote[red]{prateek}{}{[#1]}}
    
    \newcommand{\yixin}[1]{\gennote[purple]{yixin}{}{[#1]}}
    \newcommand{\grace}[1]{\gennote[magenta]{grace}{}{[#1]}}
    
    \newcommand{\cameranew}[1]{{#1}}

\else
    \newcommand{\gennote}[4][blue]{}
    \newcommand{\joel}[1]{}
    \newcommand{\liangw}[1]{}
    \newcommand{\henry}[1]{}
    \newcommand{\note}[1]{}
    \newcommand{\todo}[1]{}
    \newcommand{\adrian}[1]{}
    \newcommand{\prateek}[1]{}
    
    \newcommand{\yixin}[1]{}
    \newcommand{\grace}[1]{}
    
    \newcommand{\cameranew}[1]{{#1}}
\fi

\newcommand{\tabref}[1]{{Table~\ref{#1}}}
\newcommand{\figref}[1]{{Figure~\ref{#1}}}
\newcommand{\secref}[1]{{\S\,\ref{#1}}}

\setlist[itemize]{leftmargin=1em, itemsep=-0.1em, itemindent=0em}
\setlist[enumerate]{leftmargin=1.2em, itemindent=0em, itemsep=-0.1em}
\setlist[description]{leftmargin=1em}

\renewcommand{\paragraph}[1]{\noindent\textbf{#1.}}

\newcounter{lineno}

\PassOptionsToPackage{hyphens}{url}

\usepackage{tikz}
\usepackage{amsmath}

\usepackage{filecontents}

\usepackage{xurl}
\usepackage[english]{babel}
\usepackage{boldline} 
\usepackage{makecell}
\usepackage[linesnumbered,ruled,noend]{algorithm2e}

\begin{document}

\thispagestyle{empty}

\title{How Effective is Multiple-Vantage-Point Domain Control Validation?}

\author{
{\rm Grace Cimaszewski$^{\dag*}$}
\and
{\rm Henry Birge-Lee$^{\dag*}$}
\and
{\rm Liang Wang$^\dag$}
\and
{\rm Jennifer Rexford$^\dag$}
\and
{\rm Prateek Mittal$^\dag$}
\smallskip 
\\
$^\dag$ Princeton University  \,\,\, $^*$Joint lead author
} 

\maketitle

\begin{abstract}
Multiple-vantage-point domain control validation (multiVA) is an emerging defense for mitigating BGP hijacking attacks against certificate authorities. While the adoption of multiVA is on the rise, little work has quantified its effectiveness against BGP hijacks in the wild. We bridge the gap by presenting the first analysis framework that measures the security of a multiVA deployment under real-world network  configurations (e.g., DNS and RPKI). Our framework accurately models the attack surface of multiVA by 1) considering the attacks on DNS nameservers involved in domain validation, 2) considering deployed practical security techniques such as RPKI, 3) performing fine-grained internet-scale analysis to compute multiVA resilience (i.e., how difficult it is to launch a BGP hijack against a domain and get a bogus certificate under multiVA). We use our framework to perform a rigorous security analysis of the multiVA deployment of Let's Encrypt, using a dataset that consists of about 1 million certificates and 31 billion DNS queries collected over four months. Our analysis shows while DNS does enlarge the attack surface of multiVA, the  of Let's Encrypt's multiVA deployment still offers an 88\% median resilience against BGP hijacks, a notable improvement over 76\% offered by single-vantage-point validation. RPKI, even in its current state of partial deployment, effectively mitigates BGP attacks and improves the security of the deployment by 15\% as compared to the case without considering RPKI. Exploring 11,000 different multiVA configurations, we find that Let's Encrypt's deployment can be further enhanced to achieve a resilience of over 99\% by using a full quorum policy with only two additional vantage points in different public clouds.
\end{abstract}
\section{Introduction}

\label{sec:intro}
Certificate Authorities (a.k.a. Certification Authorities or CAs) serve as a root of trust for secure TLS communications by signing digital certificates that tie a server's identity (like a domain name) to its public key.
However, the process that CAs use to verify domain ownership, known as domain control validation, is vulnerable to BGP attacks~\cite{birge2018bamboozling, brandt2018dv++}.
By hijacking traffic to a victim's domain during domain control validation, an adversary can fool a CA into signing a certificate on behalf of the adversary for a domain the adversary does not control~\cite{birge2018bamboozling}. The first real-world examples of these attacks were observed in 2022 when adversaries used them to steal millions of dollars worth of cryptocurrency~\cite{coinbase2022celer_bridge,birgelee2022klayswap}; and millions of other websites (including those observing best security practices) are potentially vulnerable~\cite{birgelee2022klayswap}. This is also a potentially devastating attack in the hands of repressive regimes that have been accused of launching strategic BGP attacks in the past~\cite{demchak2018china}.

To mitigate the risks of BGP attack on domain control validation, several CAs (including the world largest web PKI CA Let's Encrypt and Google's CA Google Trust Services) have deployed a countermeasure known as multiVA ~\cite{birgelee2021experiences,hurst2022google_multiple_vantage_points}.\footnote{Consistent with Birge-Lee \emph{et al.} ~\cite{birgelee2021experiences}, we use the term multiVA to refer to this technology, which has also been called Multiple-Vantage-Point Domain Validation~\cite{birge2018bamboozling}, Multi-Path Domain Validation~\cite{cloudflare2019multiple_vantage_points}, and Multi-Perspective Domain Validation~\cite{lets_encrypt2020multi_perspective_validation} in some prior works.
} In multiVA, domain control validation for each certificate request is performed multiple times from different vantage points spread throughout the Internet. This technique detects BGP attacks by exploiting the fact that many BGP announcements are localized and only affect a portion of Internet traffic~\cite{birgelee2021experiences}. If a CA performs domain control validation from a vantage point whose routing is not affected by the adversary's attack, this vantage point will have its traffic routed to the victim's server and connect to the "real" domain, allowing the CA to detect the attack.

While this countermeasure is sound in principle (i.e., more remote vantage points with greater geographic spread have more potential to catch localized BGP attacks~\cite{birge2018bamboozling,birgelee2021experiences}), a key question is: \emph{how effective is multiVA in practice?} For multiVA to be effective, enough vantage points have to route to the victim during a BGP attack so that the CA can detect the attack. There is currently an active debate over multiVA's effectiveness and how to strengthen domain control validation against BGP attacks at the CA and Browser Forum (the governing body for CAs)~\cite{wilson2022protection_against_bgp_attacks}. A key challenge is that strategic BGP attacks could specifically target the weakest link in multiVA deployments, including any vulnerable IP prefixes involved in the domain validation process. 
Thus, it is not obvious that a multiVA deployment constrained by financial/operational costs and false positives can significantly reduce the BGP attack surface~\cite{brandt2021dns_resilience}. MultiVA's effectiveness in light of real-world domain hosting practices and routing dynamics is a critical and open research question for the security community~\cite{dai2021downgrade, brandt2021dns_resilience} and our work aims to inform this debate via a rigorous quantitative study. 

\begin{table*}[t]
\centering
\small
\begin{tabular}{|p{15cm}|c|}
\hline 
\multicolumn{1}{|c|}{\bf Significant Findings} & \textbf{Section} \\ \hline 
\textbf{\cameranew{Importance of DNS and RPKI}} & \\
\addtolength{\leftskip}{.2cm} - 
Considering DNS causes a five-fold increase in the number of prefixes an adversary can target & \secref{sec:le-domain-profile} \\
\addtolength{\leftskip}{.2cm} - Only 5.6\% of domains are fully  protected by DNSSEC & \secref{sec:le-domain-profile} \\
\addtolength{\leftskip}{.2cm} - 60\% of target IPs associated with domains are covered by ROAs and benefit from RPKI & \secref{sec:le-domain-profile} \\
\hline
\textbf{\cameranew{Effect on Domain Resilience}} & \\
\addtolength{\leftskip}{.2cm} - Considering the DNS attack surface drops the resilience of domains by 20\% (from 95\% to 75\%) under multiple vantage point validation but 40\%+ (from 83\% to 42\%) under single vantage point validation & \secref{sec:le-dns-analysis} \\
\addtolength{\leftskip}{.2cm} - 76.3\% of domains have some RPKI coverage of their DNS resolution graphs; current RPKI coverage improves the resilience of domains by 15\% to 90\% and full coverage could improve the resilience by 20\% to 95\% &
\secref{sec:le-dns-analysis} \\
\addtolength{\leftskip}{.2cm} - The resilience of Let's Encrypt's current deployment is still at 88.6\% considering these factors & 
\secref{sec:le-dns-analysis} \\
\hline
\textbf{Ways to Improve the Deployment} & \\
\addtolength{\leftskip}{.2cm} - Adding a single vantage point in Let's Encrypt's current cloud provider improves resilience to 93.2\%
 & \secref{sec:le-res-improve} \\
\addtolength{\leftskip}{.2cm} - If two vantage points are added, a cross-cloud strategy is optimal and can boost resilience to 97.5\% & \secref{sec:le-res-improve} \\
\addtolength{\leftskip}{.2cm} - Implementing a full quorum policy even with existing vantage points can boost resilience to 96.5\% & \secref{sec:le-res-improve} \\
\hline
\end{tabular}
\caption{Significant results from our analysis framework.} 
\label{tab:lessons}
\end{table*}

There are several components of the Internet and routing ecosystem that impact the effectiveness of multiVA, including the Domain Name System (DNS)~\cite{dai2021downgrade,brandt2021dns_resilience} and the ongoing deployment of the Resource Public Key Infrastructure (RPKI). While previous work~\cite{brandt2021dns_resilience} has only begun to address some of these factors (as we discuss in \secref{sec:related:domain_validation}), we present the first work to rigorously compute the resilience of multiVA against BGP attacks under real-world network and DNS dynamics.

\bheading{Contributions.} In this paper, we contribute a novel analysis framework that models multiVA under real-world conditions to rigorously understand its effectiveness in the wild. Notably, our work models the DNS infrastructure associated with customer domains of Let's Encrypt's real deployment (at the scale of over a million domains) and incorporates the existing RPKI deployment via Internet-scale topology simulations. We also present the first analysis of vantage points deployed across different cloud providers.
\begin{enumerate}
    \item Using log data from the world's largest web-PKI CA Let's Encrypt, we perform an extensive study of the DNS configurations of over a million customer domains using \textbf{31 billion DNS queries} from geographically-distributed vantage points. We develop a custom full-graph DNS resolver to capture factors that impact the resilience of domains to BGP attacks, including IP prefixes of associated DNS infrastructure, and capturing the deployment of DNSSEC and RPKI.
    \item We perform fine-grained Internet topology simulations under both current and future RPKI conditions using CAIDA Internet topology data~\cite{caida}, BGP data collected by public route monitors, and traceroute data from multiple cloud providers to every prefix on the Internet to understand the routing of a diverse set of 19 cloud datacenters during attacks from 1\,K sampled adversaries against all 800\,K IP prefixes on the Internet (which required four hundred million full Internet topology simulations containing 24 trillion simulated routes).
    \item We performed a quantitative multi-faceted security evaluation to measure the impact of factors in the routing and DNS ecosystems on over 11\,K different hypothetical simulated multiVA deployments (including deployments using multiple cloud providers), as well as Let's Encrypt's existing deployment. By computing effective resilience\footnote{The fraction of ASes on the Internet that cannot use an equally-specific BGP attack to obtain a certificate for a given domain~\cite{birge2018bamboozling,birgelee2021experiences}.} for Let's Encrypt customer domains under different conditions, we determine the viability of multiVA and understand how to optimize a multiVA deployment.
\end{enumerate}

The results of this analysis (also summarized in Table~\ref{tab:lessons}) show several insights to inform multiVA deployments: 
\begin{enumerate}
    \item Considering DNS as an attack surface multiplies the number of IP prefixes an adversary can target with a BGP attack by a factor of five. Only 5\% of domains are protected by DNSSEC making the DNS attack highly viable.
    \item 60\% of target IPs are covered by RPKI Route Origin Authorizations (ROAs) which helps to greatly reduce the attack surface introduced by DNS.
    \item Considering both the DNS and the RPKI infrastructures,  the resilience of the median customer domain under Let's Encrypt's current multi-VA deployment is still 88.6\% (only 6\% less than when both of these factors are omitted). 
    \item Let's Encrypt's deployment can be further strengthened by adding additional cross-cloud provider vantage points or by
    switching to a full quorum policy (where successful validation at all vantage points is required). These changes can boost the resilience of the median domain to above 98.6\% or 96.5\% respectively. Both of these changes together can yield a 99.3\% median domain resilience. 
\end{enumerate}

Our findings present critical empirical evidence for promoting the adoption of multiVA. Our analysis framework that accounts for the attack surface introduced by DNS, as well as protections offered by RPKI, has broad applicability to other network security and privacy domains beyond the PKI. We open source our framework to facilitate CAs and future research. 

\bheading{Ethical considerations.}
Our analysis uses the certificate issuance logs from Let's Encrypt as a feed. The logs have been sanitized to remove any sensitive client information. Moreover, all the information we use is publicly available in certificate transparency logs. In our measurement, we rate limit our queries to avoid overwhelming the public DNS resolvers.

\section{Background}

In this section we start with a brief background on interdomain routing insecurities and then provide an overview of the DNS and how it is susceptible to routing attacks.

\subsection{Routing System: BGP Attacks}
Interdomain routes between Autonomous Systems (ASes) are negotiated via the Border Gateway Protocol (BGP). When an AS announce its IP prefix via BGP and lists its Autonomous System Number (ASN) as the origin of that BGP announcement, neighboring ASes redistribute that announcement and append their own ASNs to it (providing a list of ASes traffic will traverse to reach the origin AS in each announcement). However, BGP route messages are unsigned and unauthenticated and thus vulnerable to BGP hijacks. The frequency, stealthiness, and financial damages associated with BGP hijacks are on the rise (as seen in the repeated strategic use of BGP attacks to steal cryptocurrency~\cite{birgelee2022klayswap,coinbase2022celer_bridge}). 

\bheading{Equally-specific prefix hijacks.}
One of the simplest forms of BGP attacks is an equally-specific prefix hijack where an adversary makes a BGP announcement containing the exact same prefix belonging to a victim AS, in effect claiming it is the owner of its victim's IP prefix. In this attack, affected ASes believe this malicious announcement and route their traffic destined for the victim to the adversary instead. When undetected, this is largely effective at attracting traffic.

\bheading{Sub-prefix hijacks.}
In sub-prefix (or more-specific) prefix hijacks, the adversary announces a sub-prefix (i.e., a longer and more preferred prefix) of the victim's IP prefix. Because of longest-prefix-match forwarding, the adversary's sub-prefix route is preferred over the victim's route, often enabling these attacks to affect the entire Internet. While sub-prefix attacks are highly effective, they are not always viable as ASes typically filter BGP announcements for prefixes longer than 24 bits (or 48 bits in IPv6). Thus, 24-bit IPv4 network prefixes usually cannot be attacked with sub-prefix hijacks.

\bheading{RPKI makes BGP hijacks more difficult.} 
Resource Public Key Infrastructure (RPKI)~\cite{rpki} cryptographically attests to which ASes own which IP prefixes. Thus, when adversaries illegitimately try to originate an IP prefix which was not allocated to them, RPKI-based filtering (known as Route Origin Validation or ROV) can block these announcements and prevent them from propagating. Additionally, RPKI also specifies the prefix length that should be used in BGP announcements for that prefix. This allows ASes performing ROV to also filter all sub-prefix attacks (independent of which ASN is listed as the origin).

While RPKI is a major improvement to routing security, it is not a panacea particularly because it only validates the origin AS listed in a BGP announcement and not the other ASes that claim to forward traffic to the origin. This allows an adversary to, instead of claiming to originate a victim's IP prefix, maliciously claim a (potentially non-existent) route to the true origin AS in a BGP announcement. We refer to malicious announcements of this type as equally-specific-prefix prepend attacks and these attacks can evade ROV but tend to affect a smaller portion of the Internet (as this strategy makes the adversary's announcement less preferable in BGP route selection).

\subsection{DNS Name Resolution}
\label{sec:background-dns}

When a web client connects to a domain, the domain name must be resolved to an IP address which hosts that domain. This is done using the Domain Name System or DNS. The DNS is hierarchically composed of delegations of zones between nameservers. When a client resolves a domain, the client first queries the root nameserver, then follows the graph of nameserver delegations to reach the authoritative nameserver for that domain, which provides the actual record containing the domain name's IP address.

Most DNS queries are sent over unencrypted, unauthenticated UDP packets making them a target for network attacks. In addition to off-path vulnerabilities (like packet fragmentation-based attacks~\cite{dnsfragmentation}), DNS is vulnerable to BGP attacks which allow adversaries to answer DNS queries with malicious records that can point users to adversary-controlled servers (instead of a victim domain)~\cite{myetherwallet}.

\bheading{DNSSEC adds cryptographic protections to DNS.} DNSSEC is an extension to DNS that requires all records from authoritative nameservers to be cryptographically signed. DNSSEC prevents attacks on DNS because even if an adversary is able to compromise a victim's DNS query, it cannot generate a valid, signed DNS response without also having access to the victim's private DNSSEC key. While many CAs perform DNSSEC validation when resolving domain names, not all domains participate in DNSSEC.

\section{Adversary Model}

Our threat model is an extended version of the original threat model used in the design of multiVA~\cite{birgelee2021experiences} that accounts for the DNS attack surface. The goal of the adversary is to obtain a certificate for a domain it does not control. This certificate can then be used to impersonate the domain or decrypt TLS traffic between the victim domain and domain visitors in a man-in-the-middle attack.

\bheading{Trusted system components.}
We assume that the CA and all of its remote vantage points are trusted (the CA is not acting maliciously and it has complete control over its remote vantage points). We do not consider non-BGP attacks in our threat model like attacks that exploit off-path vulnerabilities in the DNS protocol~\cite{kaminsky2018dns, brandt2018dv++}. Finally, we do not consider vulnerabilities introduced by bugs/misconfigurations in the software run by the CA or its vantage points. 

\begin{figure}[t]
    \centering
    \includegraphics[width=0.8\linewidth]{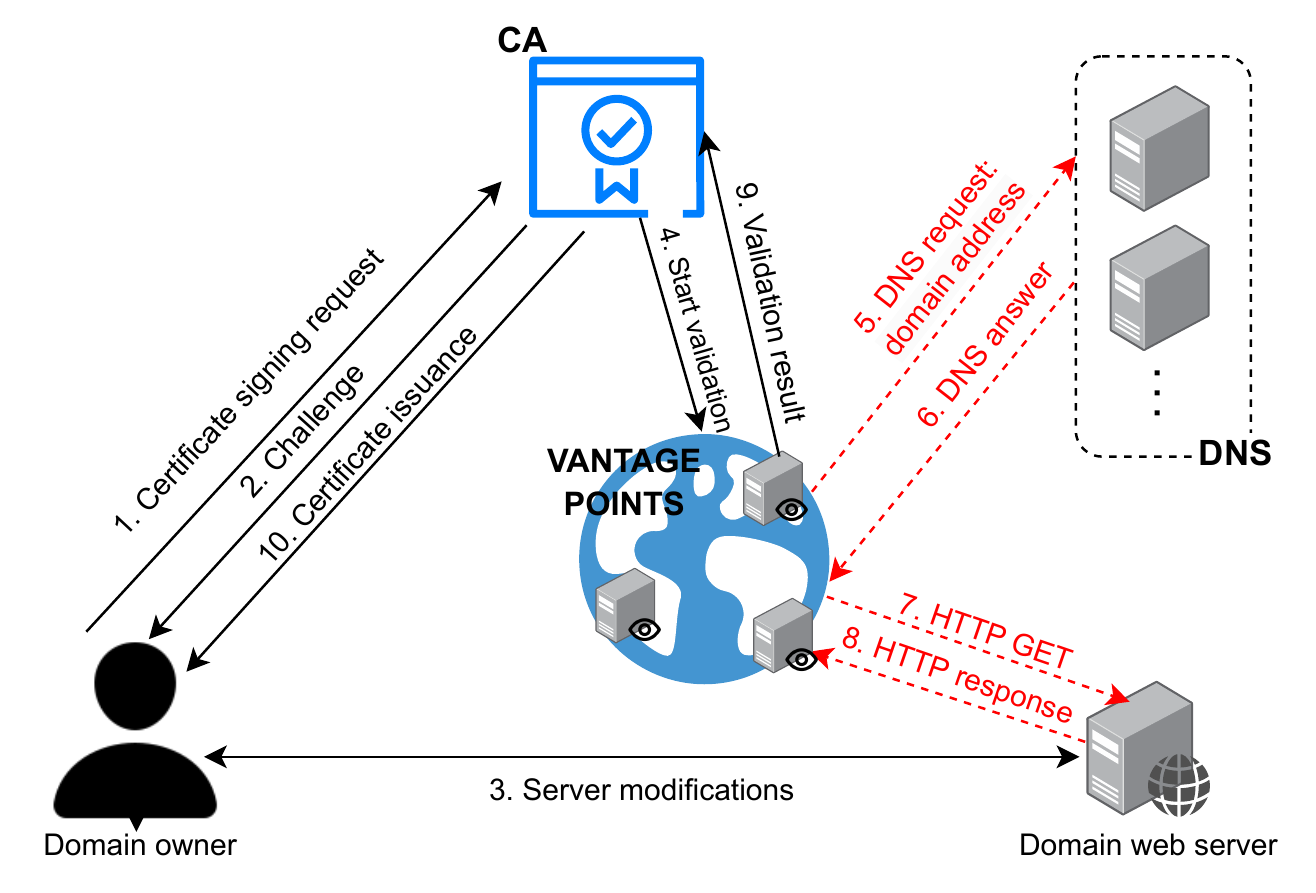}
    \caption{Certificate issuance process under multiVA. (Red text and lines indicate connections vulnerable to BGP hijack.)}
    \label{fig:cert-issuance-flow}
\end{figure}

\begin{figure}[t]
     \centering
    \includegraphics[trim=2cm 0 3cm 0, clip, width=0.9\columnwidth]{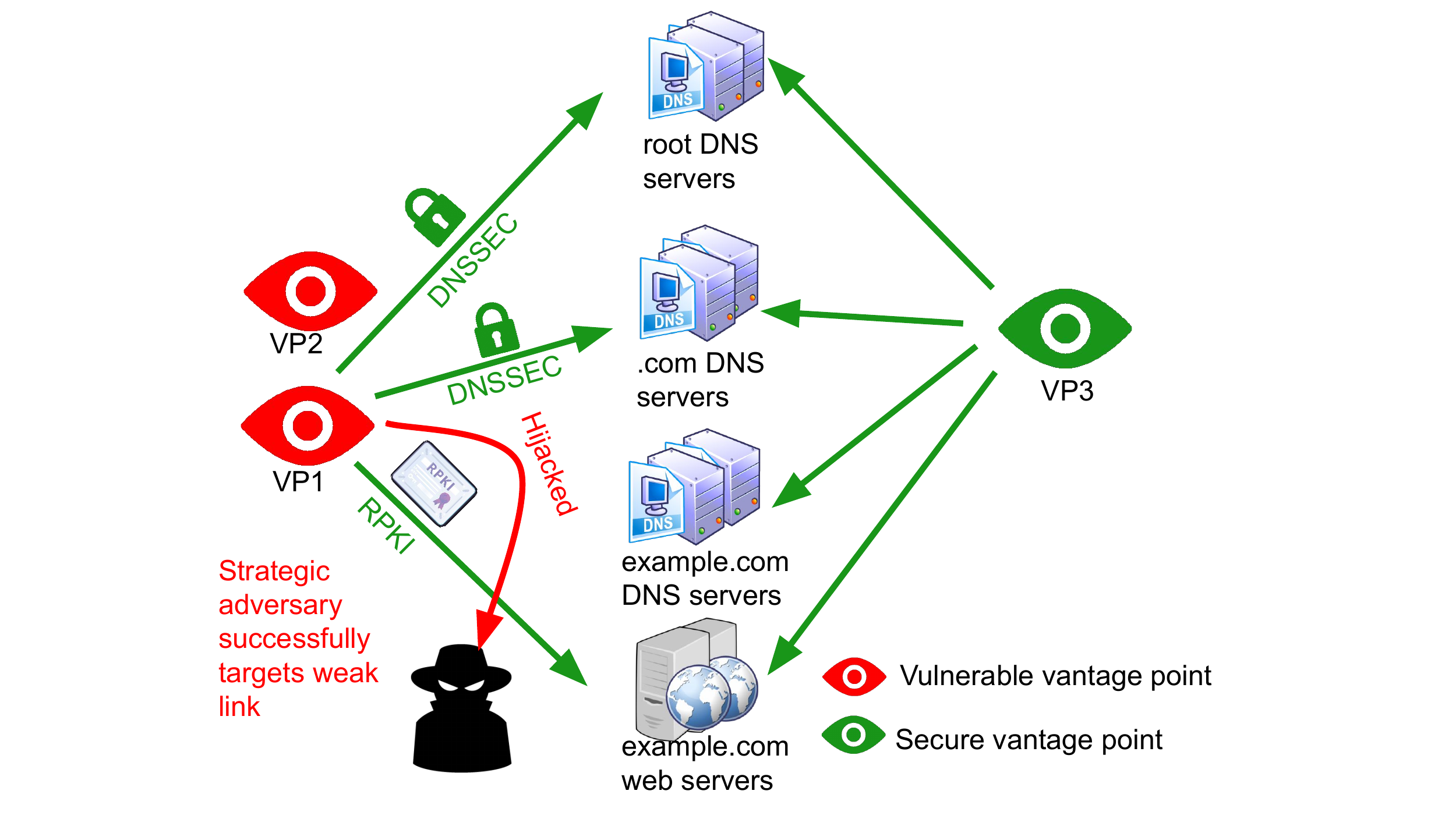}
    \caption{Example of vantage points and a domain vulnerable to an adversary's BGP hijack. The adversary strategically picks the most vulnerable link to hijack in light of all countermeasures like RPKI and DNSSEC. In order to obtain a malicious certificate for the target domain, the adversary must hijack a sufficient number of vantage points to satisfy the CA's quorum policy (2 of 3 in our example). 
    }
     \label{fig:dns_attacks}
\end{figure}

\bheading{Adversary attack strategy.}
We consider an adversary that aims to use a BGP attack to fool the domain control validation process. Because domain control validation is used to bootstrap trust for the first time, domain control validation has to be performed over unauthenticated channels. By launching a BGP attack against a victim domain that affects domain control validation traffic from a CA and sufficiently many remote vantage points, an adversary can pose as the victim domain and obtain a malicious certificate (see \figref{fig:dns_attacks}).

\bheading{BGP capabilities.}
We consider an adversary with control of a single malicious AS that can make a malicious BGP announcement for any prefix (or prefixes) it chooses to target. We consider that the adversary may launch the following two types of BGP attacks:
\begin{enumerate}
    \item \textbf{Equally-specific-prefix attack:} an adversary announces an equal-length prefix as the victim domain's prefix.
    \item \textbf{Equally-specific-prefix prepend attack:} an adversary claims reachability to the victim's prefix via a non-existent connection by inserting itself on a valid path. This attack attempts to circumvent RPKI-based ROV detection by keeping the valid originating AS as the route announcement origin.
\end{enumerate}

We focus our analysis on equally-specific prefix attacks and omit sub-prefix hijacks from our analysis because this is the primary threat model motivating the design of multiVA~\cite{birge2018bamboozling, birgelee2021experiences}. This is justified as many prefixes on the Internet are not vulnerable to sub-prefix attacks. RPKI can prevent sub-prefix hijacks if a given prefix registers an ROA record with appropriate max-length attribute set~\cite{RPKI_max_length} and its BGP neighbors perform ROV. Let's Encrypt's vantage remote points (and all additional vantage points we considered) perform ROV ~\cite{awsrov,gcprov,azurerov} and many top content networks hosting domains (e.g., Google, Amazon, CloudFlare, and Microsoft) have adopted RPKI~\cite{isbgpsafeyet}. Furthermore, RPKI adoption is steadily increasing. We found that 60\% of IP addresses associated with TLS domains were protected by RPKI (see \tabref{tab:dns-profile}). In addition, IP prefixes announced with the maximum length for interdomain routing (24 bits for IPv4 and 48 bits for IPv6) are immune to sub-prefix attacks as well and this has been propsed as a security measure for high-security services~\cite{sun2015raptor}. Our study found that $89.5\%$ of target prefixes were either /24 or had valid ROA records registered, suggesting that subprefix hijacks present limited opportunities to an attacker in the context of multiVA. Furthermore, sub-prefix attacks are more easily identified with network monitoring, making network-layer defenses (e.g., ARTEMIS~\cite{sermpezis2018artemis}) more viable.

\bheading{Incorporating the DNS attack surface.}
We extend the original multiVA threat model to consider an adversary can attack the DNS infrastructure associated with a victim's domain (via a BGP attack) in addition to server A record IP addresses (see \figref{fig:dns_attacks}). When a CA performs domain control validation, before contacting the domain to complete validation, the CA must resolve the domain to an IP address via the DNS. If an adversary hijacks a vulnerable DNS query with a BGP hijack, the adversary can generate a malicious DNS response that directs the CA to an adversary-controlled server (instead of the victim domain) to fraudulently complete validation.

This significantly expands the multiVA attack surface as the many DNS nameservers contacted to resolve a domain name become attack targets. We also extend the original threat model to assume an adversary can retry validation several times to allow it to fool validation even if it can only compromise one of several A record or nameserver IP addresses. We consider all authoritative nameservers involved in the DNS lookup of a victim's domain with the exception of those protected by DNSSEC. This is because  an adversary cannot forge DNSSEC-protected DNS responses with a BGP hijack.

We exclude off-path attacks on DNS (like transaction id guessing~\cite{kaminsky2018dns} and packet-fragmentation attacks~\cite{dnsfragmentation}) from our threat model as these attacks can be mitigated with changes in a CA's DNS operations~\cite{kaminsky2018dns, dnsfragmentation} and do not require mutliVA.
\section{Analysis Framework}
\label{sec:setup}

Prior work on the security of multi-vantage-point domain validation (multiVA) does not, or only to a limited extent, consider DNS as a potential attack vector~\cite{birge2018bamboozling, birgelee2021experiences,brandt2021dns_resilience}. Our study aims to understand the security of multiVA under a more realistic setting, which not only considers additional DNS-based attack vectors, but also the relevant deployed security measures (e.g., RPKI and DNSSEC) and operational practices.  The former may allow more attack strategies and degrade the security of multiVA, while the latter could improve security. Toward this goal, we design a new analysis framework to facilitate our security evaluation, as the methods used in prior work are not sufficient. We develop our analysis framework as a general-purpose, automated tool so that any CA can use our framework to evaluate the security of its multiVA deployment against routing attacks. Furthermore, our framework is capable of modeling the current setup as well as other potential deployment options, which can help a CA to optimize the security of its multiVA deployment.      

Our framework takes as input a set of domain names and configuration of a multiVA deployment, and outputs the domain's resilience. A domain's resilience measures the fraction of the Internet from which that domain is \emph{immune} to equally-specific BGP attacks. ASes in this fraction of the Internet cannot fool enough CA vantage points with an equally-specific BGP attack for the CA to potentially sign a malicious certificate for that domain. We  measure the impact of various multiVA configurations by seeing how the resilience of customer domains signed by the CA changes. 
We defer the full definition of resilience to \secref{sec:resilience}.

\begin{figure}[t]
\centering
\includegraphics[trim=1.8cm 6.0cm 1.8cm 0.8cm, clip, width=\columnwidth]{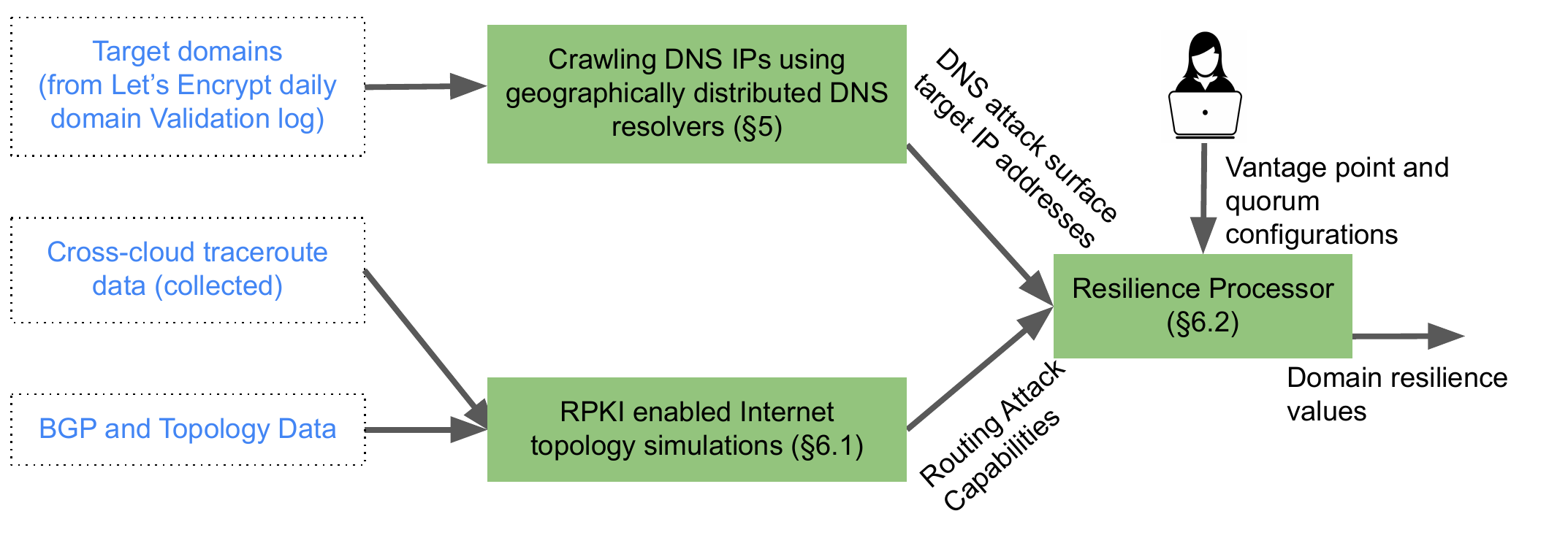}
\caption{An overview of the analysis framework.}
\label{fig:analysis_framework}
\end{figure}

As shown in \figref{fig:analysis_framework}, our framework consists of three major components: (1) the geographically-distributed DNS resolvers, (2) the Internet topology simulator, and (3) the resilience processor.

\bheading{Geographically-distributed DNS resolvers.}
To get a more complete view of the potential DNS attack surface, the \emph{geographically-distributed DNS resolvers} resolve the input domains (extracted from Let's Encrypt's  certificate issuance logs in our work) in near-real-time from geographically distributed locations, and trace the dependencies of the resolved domains (i.e., full-graph DNS resolution). Over the course of our experiment (which studied approximately 1.4 million domains) we sent 30.7 billion DNS requests and recorded 520 billion nameserver IP addresses and 30.6 billion A record IP addresses. We use the resulting data along with BGP hijack simulations to understand which IP prefixes (and which domains) are potentially vulnerable to BGP attacks. The domain resolution routine is executed each time a daily issuance log is received from Let's Encrypt to allow for domains to be resolved within 24 hours and reduce the impact of domains being taken down or changing DNS configurations.

\bheading{Fine-grained topology and accurate BGP attacks simulation incorporating RPKI.}
To produce accurate BGP attack simulations, the \emph{Internet topology simulator} performs fine-grained, prefix-level (instead of AS-level) routing simulation.

To better model the routing decisions of the datacenters we wanted to consider for vantage points, we needed to have an accurate understanding of the BGP connectivity of these datacenters. We used the bdrmap tool~\cite{bdrmap} to run traceroutes to every prefix on the Internet and infer peer lists for 19 different datacenters spread across AWS, Azure and GCP (see Appendix \tabref{tab:cloud-prov-location} for a full list). This ensured we could produce accurate simulations of these datacenters and appropriately understand how they would be affected by BGP hijacks.

Leveraging this traceroute data, public topology information~\cite{caida}, and BGP RIB dumps~\cite{ripe_ncc_ris,routeviewUO}, the simulation engine simulates the interdomain routing of roughly two hundred thousand groups of IP prefixes, and further simulates equally-specific prefix BGP attacks against these prefixes launched from 1,000 random ASes. Different from prior BGP attack simulations~\cite{birge2021le}, our simulation engine takes deployed BGP security practices, i.e., RPKI, into account, and can simulate attacks against RPKI-protected prefixes. Thus, these simulations are repeated under both RPKI and non-RPKI conditions resulting in roughly four hundred million simulations and over twenty-four trillion AS-level paths calculated. 

\bheading{Multifaceted quantitative security estimation.}
Finally, using the DNS data collected by the measurement engine and the results produced by the simulation engine, the \emph{resilience processor} estimates the security of a multiVA deployment against routing attacks, which could target any vulnerable DNS servers found by the geographically-distributed DNs resolvers, under various scenarios.

Using the security estimation result as an indicator, our research addresses the following real-world deployment questions for CAs to maximize their resilience to BGP hijacks:
\begin{itemize}
    \item \textbf{Selection of vantage point location.} What are the optimal vantage point locations that make the CA most secure against BGP attacks?
    
    \item \textbf{Measuring impact of quorum policy.} Birge-Lee \emph{et al.}~\cite{birgelee2021experiences} observed that a stricter quorum policy may have negative operational implications for a CA's issuance (e.g., benign failures). By what extent do stricter quorum policy configurations improve resilience against BGP hijacks?
    
    \item \textbf{Comparison of cloud provider routing resiliency.} Does validation using vantage points hosted in multiple cloud providers provide additional security benefits (as opposed to concentrating VPs in one cloud provider)?
\end{itemize}
 
\section{Measuring the DNS Attack Surface}
\label{sec:dns}
In order to calculate domain resilience, we develop a global experimental system to perform DNS lookups of domains in certificates signed by Let's Encrypt. The design of our DNS resolver tool was driven by our primary research motivation: \textbf{to quantify the extent of the DNS attack surface that could be the target of interdomain routing attacks}. Our tool collects comprehensive details of the DNS lookup graph for each domain, \textit{to record every IP address that may be contacted in the resolution of that name.}

\begin{figure}[t]
\centering
\includegraphics[trim=0 1.5cm 0 0, clip, width=0.9\linewidth]{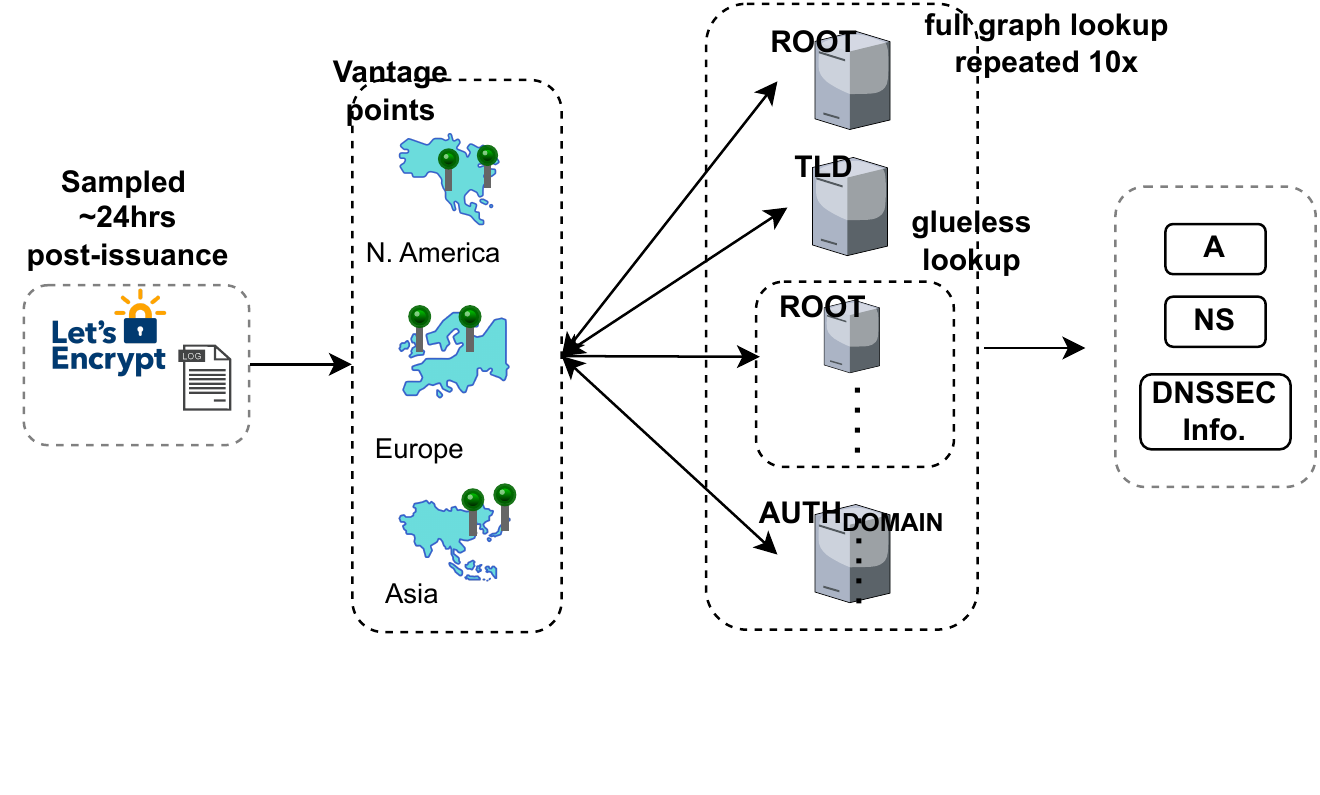}
\caption{Design of daily DNS resolution framework, using 6 distinct vantage points across 3 geographic continents.}
\label{fig:dns_framework}
\end{figure}

\subsection{Defining the DNS Attack Surface}
\label{sec:dns-define}

We denote the \emph{full-graph DNS attack surface} of a domain $d$ as $Q(d)$, which is the set of all IP addresses that may be the targets of BGP hijacks for $d$. Intuitively, $Q(d)$ includes all IP addresses of the webservers that host $d$, as well as all IP addresses that may be contacted to resolve the DNS zone that hosts $d$'s DNS records. To illustrate our mathematical formulation, we will refer to \texttt{example.com} as a running example.

Define the DNS zone delegation chain of $d$ as $Z_0, Z_1, \ldots, Z_{k_d}$, where $Z_0$ represents the root zone and $Z_{k_d}$ the authoritative zone for $d$'s DNS records. Each zone $Z_i$ is hosted by a set of nameservers $N_{Z_i}$ ($n \in N_{Z_i}$ indicates a nameserver name). Lastly, denote the set of IP addresses contained in the A record of a name $d$ as $A(d)$.  To resolve domain $d$'s IP address $A(d)$, a DNS resolver iteratively queries nameservers in each successive zone, receiving either a delegation to the next zone or an answer (or error) from the authoritative nameserver(s) $N_{Z_{k_d}}$. We use $I(Z)$ to denote the IP addresses of the nameservers in zone $Z$, as well as any IPs that may be queried to resolve the namerserver names themselves.

For the case of $d = \texttt{example.com}$, \texttt{example.com} is delegated across DNS zones $Z_0 = .$, $Z_1 = \text{.com}$, and $Z_2 = \text{example.com}$. $N_{Z_0}$ consists of the 13 IANA-defined root nameservers; $N_{Z_1}$ consists of the .com TLD authoritative nameservers; $N_{Z_2} = \{\texttt{ns1.provider.net},\, \texttt{ns2.provider.net}\}$ consists of the nameservers authoritative for example.com. 

We formalize the attack surface of $d$, $Q(d)$, as follows:
\begin{equation*}
Q(d) = I(Z_{k_d}) \cup A(d)
\end{equation*}

The DNS resolver can learn the IP address of $d$'s nameservers by two methods: either by sending a separate DNS request for the nameserver's A records with its own associated $Q(n)$, or using the ``glue'' records provided in the additional section of the parent zone's response (if included). We refine the expression of $Q$ for this case and call this $Q^{\prime}$:
\begin{equation*}
    Q^{\prime}(n) = 
    \begin{cases} 
    A(n) & \text{if glued record exists} \\
    Q(n) & \text{otherwise (``glueless'')} \\
    \end{cases}
\end{equation*}

The DNS attack surface of $d$'s authoritative zone $Z_{k_d}$ can be defined recursively: 
\begin{equation*}
    I(Z_{k_d}) = I(Z_{{k_d}-1}) \cup \bigcup\limits_{n \in N_{Z_{k_d}} } Q'(n)
\end{equation*}
Applying this algorithm to compute $Q(\texttt{example.com})$, we must resolve the A records of \texttt{example.com} as well as enumerate all IP addresses needed to resolve its authoritative nameservers $\{\texttt{ns1.provider.net},\, \texttt{ns2.provider.net}\}$. To do this, the resolver sends a DNS query to the root nameservers, which responds with a delegation to the .com nameservers in the authority section and a list of these nameserver IPs in the additional section. Next, the query is sent to the .com nameservers, which respond with delegation to the $\{\texttt{ns1.provider.net},\, \texttt{ns2.provider.net}\}$ authoritative nameservers. Separate DNS queries for the A records of both $\texttt{ns1.provider.net}$ and $\texttt{ns2.provider.net}$ are sent to resolve the authoritative nameservers' IP addresses. Lastly, using the answer from this ``glueless'' lookup, the authoritative nameservers are queried for the A records of \texttt{example.com}.  

Next, we consider DNSSEC records. We refine this expression for the case where the preceding zone's response contains a DS signing key, if DNSSEC records are registered: 

\begin{equation*}
    \mathcal{I}^{+DNSSEC}(Z_i) = 
    \begin{cases}
        \emptyset & \text{if DS key in $Z_{i-1}$ response} \\
        I(Z_i) & \text{otherwise} \\
    \end{cases}
\end{equation*}

For the base case of the root zone $Z_0$, the full-graph DNS attack surface is the empty set: the IP addresses of the root nameservers are hardcoded in operating system root hints files and the root nameserver responses are signed by the global DNS root key, so the root zone is safe from BGP hijacks:
\begin{equation*}
    I^{+DNSSEC}(Z_0) = \emptyset
\end{equation*}

Combining these equations, the quantity $Q^{+}(d)$ represents the set of IP addresses that may be BGP hijacked when DNS resolving and connecting to the domain $d$. $Q^{+}(d)$ can be derived by performing an iterative DNS query starting at the root nameservers and resolving/recording the glued IP addresses of \emph{every nameserver} in each successive zone.

\begin{equation*}
    Q^{+}(d) = I^{+DNSSEC}(Z_{k_d}) \cup A(d)
\end{equation*}

\subsection{Designing a Full-Graph DNS Resolver}
\label{sec:dns-tool}

While prior work had developed the notion of full-graph DNS and its implications for domain name hijacking, this work focused on DNS-level exploits and did not consider the effect of insecure Internet routing \cite{ram2005transtrust}. There has also been little exploration of the usage of DNSSEC and its implication within the context of the PKI and routing security. (A more thorough comparison of prior work can be found in \secref{sec:related}.)

Our lookup tool aimed to overcome several limitations in previous DNS resolver work, chiefly: 1) the lack of full-graph resolution in open-source tools like KnotDNS or Unbound; 2) including record and IPv4/v6 address dependencies in the DNS trust graph, beyond simple nameserver delegation chains as in \cite{ram2005transtrust}; 3) the omission of DNSSEC records in calculating the attack surface.

Our DNS resolver tool differs from classic open-source resolver implementations and prior research works in the following aspects: 

\bheading{Full graph DNS lookups considering DNSSEC.} Referring to the definition in \secref{sec:dns-define}, the resolver tool computes the full-graph DNS attack surface $Q^{+}(d)$ of the domain name $d$, which is a set of IP addresses including $d$'s A records and all its nameserver dependencies. The tool also records the presence of DNSSEC DS records in each successive zone response, as the DS key provides a cryptographic authenticity check on the DNS response that can be validated by Let's Encrypt and other certificate authorities. DNS responses from the root zone and many top-level domain zones (as well as the authoritative resolver on about 5.6\% of domains) have DS key records securing their responses.

\bheading{DNS lookups near realtime to cert issuance.} Because TLS certificates do not certify fields related to routing or DNS, a domain's DNS configuration and records are subject to change any time after certificate issuance. Previous work found that DNS queries for a high proportion of domains certified by Let's Encrypt returned NXDOMAIN errors when queried weeks after certificate issuance time~\cite{birge2021le}. To mitigate these effects, we perform DNS lookups on domains shortly after issuance (i.e., less than 24h) to ensure DNS conditions that closely matches the one performed by Let's Encrypt.

\bheading{Distribution DNS lookups geographically.} A domain name may resolve to different IP addresses when queried from different geographic regions. This behavior is indicative of Content Distribution Network (CDN) use, and we found that connections to local content replicas are often much shorter than the connection to a single origin server potentially in another region, which  increases resilience to hijacking.

\bheading{Repeated DNS lookups.} We observed that around 3\% of domains resolved to different IP addresses when a DNS lookup was performed multiple times. The two major factors influencing this were 1) DNS load-balancing systems where only a random subset of IP addresses (drawn from a large pool of replicas) were returned on each query and 2) path-dependent domains where queries to different authoritative nameservers returned answers with different A records. Since an adversary can realistically re-attempt obtaining a certificate several times (and even impact which nameserver is used by the CA),  we consider any IP address that could be returned in a lookup part of the BGP attack surface and performed each lookup 10 times to discover as many IP addresses as possible.\footnote{After 10 lookups, there was a significant reduction in the number of new A record IP addresses returned, making 10 a logical cap for repeated queries.}

\subsection{Global Near-Real-Time Domain Resolution Infrastructure}

In support of our work, Let's Encrypt provided us access to daily certificate issuance and domain validation logs, which contained information for all certs issued in the preceding 24 hours (an average of 1.08 million certs per day). 
Over the period of April 20 - August 13 2022, we randomly sampled 10K certificates daily from these logs, yielding a total dataset of 1.354 million domains from 810,000 certificates. Daily DNS resolution jobs using the tool described in \secref{sec:dns-tool} were run in three continents at six AWS EC2 server locations \footnote{Let's Encrypt's current multi-VA infrastructure includes vantage points in Ohio, Oregon, and Frankfurt AWS datacenters. The AWS region codes of city names are listed in \tabref{tab:cloud-prov-location}.}: (1) \textbf{North America:} Ohio, Oregon; (2) \textbf{Europe:} Paris, Frankfurt; (3) \textbf{Asia:} Singapore, Tokyo, as close to certificate issuance time as possible (5 hours after log upload). Chiefly, we are interested in performing iterative queries
for the A (IPv4 address) and AAAA (IPv6 address) records of
domain names listed in the subject names of certificates issued
by Let’s Encrypt.

\textit{Implementation details.}
The primary implementation challenge lay in operation at scale: our tool needed to operate efficiently to conduct millions of DNS requests per day, tolerating various misinformed records and edge cases, and output a concise log that captured the full history of how the lookup was performed and that could be plugged into our analysis engine (see \figref{fig:analysis_framework}). We release both our DNS resolver tool as open-source software on our \href{https://github.com/inspire-group/routing-aware-dns}{Github repository}\footnote{https://github.com/inspire-group/routing-aware-dns} and collected DNS dataset to the public. Our implementation is around 2250 lines of Python source code and can be extended to log other records of a name's DNS graph as well.

\subsection{Profile of LE-certified Domains}
\label{sec:le-domain-profile}

\begin{table}[t]
\footnotesize
\centering
\begin{tabular}{|r|c|}
\hline 
\multicolumn{1}{|c|}{\bf Feature} & \multicolumn{1}{|c|}{\bf Figure}\\ \hline
\multicolumn{1}{|l|}{\textbf{Total number of certificates}} & 810,000\\
Number of certs. successfully resolved & 755,942\\
\hline
\multicolumn{1}{|l|}{\textbf{Total number of domains}} & 1,354,318\\
\% successful A record resolution & 97.3 \\
\% successful AAAA record resolution & 12.3 \\
\hline
\multicolumn{1}{|l|}{\textbf{IP prefixes of domains}} & \\
Median number of prefixes in A records & 1 \\
Median number of prefixes in NS records & 3 \\
\hline
\multicolumn{1}{|l|}{\textbf{Little use of DNSSEC}} & \\
\% domains full DNSSEC-signed & 5.6  \\
\hline
\multicolumn{1}{|l|}{\textbf{Registration of RPKI-ROA records}} & \\
\% domains with at least 1 ROA-covered prefix & 76.3 \\
\% domains with all ROA-covered prefixes & 26.2 \\
\% target IPs with ROA records & 60.0 \\
\hline
\end{tabular}
\caption{An overview of DNS dataset, summarized.
}
\label{tab:dns-profile}
\end{table}

In this section, we present several key statistics on the routing implications of domains' DNS and webserver configurations. This represents the first study to date detailing features of domains included in Let's Encrypt certificates. Our key findings are summarized in \tabref{tab:dns-profile}.

\textit{Multiple prefixes for hijack targets.} Of the 1.248 million domain names sampled, 97.8\% and 12.3\% could be resolved to valid A and AAAA records successfully. Note that the low rate of AAAA record retrieval is not a consequence of the tool itself, but because of low IPv6 usage by the domain names surveyed. On average, the domains surveyed had webservers hosted across 1.158 distinct IP prefixes (median 1.0) and DNS nameservers hosted across 4.72 prefixes (median 3.0). 86.2\% (11.0\%), 11.7\% (29.1\%), 0.73\% (9.4\%), 0.56\% (11.1\%), 0.18\% (3.5\%) of webservers (DNS nameservers in parentheses) are associated with 1 to 5 IP prefixes, respectively. A detailed distribution is in Appendix \figref{fig:prfx-distr}. Overall, DNS presents nearly 5x more potential routing hijack targets to a hijacker than the website servers alone.
$41.9\%$ of nameservers used are hosted on /24 prefixes, meaning they are resistant to sub-prefix hijacks. Webservers tended to be hosted on shorter prefix lengths: only $35.0\%$ of webserver IPs are hosted on /24 prefixes. While we leave further analysis of sub-prefix hijacks on DNS for future work, usage of a prefix length of 24 bits and the maxLength attribute in RPKI ROAs can improve resilience against such attacks.

\textit{Sparse adoption of DNSSEC.}
We find that DNSSEC registration is sparse: only $5.6\%$ of A records returned were fully signed by DNSSEC at every link in the nameserver delegation chain, AAAA records were somewhat more fully signed at $16.0\%$. Overall, DNSSEC ensures integrity of DNS records for only a very small minority of domains. 

\textit{Growth of RPKI-ROV.} Our study gives encouraging evidence for increasing adoption of ROA registration, especially at the nameserver level. $63.6\%$ of all nameservers surveyed are hosted on IP prefixes with valid ROAs. ROA registration is slightly lower for webserver IPs, with $51.8\%$ of webserver IPs having ROAs. 

\textit{DNS nameserver centralization.} Our measurement shows the global DNS infrastructure is highly centralized in terms of prefix space and operational ownership. Only 100 prefixes, owned by 20 distinct organizations, represented 56.1\% of non-DNSSEC nameservers used across the domain dataset: Cloudflare alone operates over 24\% of these nameservers. Five Cloudflare-hosted prefixes alone accounted for 11.9\% of all nameserver prefixes surveyed. The top nameserver hosting providers are summarized in \tabref{table:ns-prov-distr}. One advantage of this centralization is that it facilitates ROA adoption as a sizeable portion of the DNS ecosystem can be protected with ROA deployment at only a handful of top providers. As shown in ~\tabref{table:ns-prov-distr}, the ROA coverage percentage of the different DNS providers is often substantially higher than the overall routing table average of 41\%~\cite{rpkimonitor} (which is in part why ROA helps to offset the attack surface introduced by DNS).


\begin{table}[t]
\centering
\footnotesize
\begin{tabular}{|c|c|c|}
\hline
\bf{Provider}        & \bf{Prop. of NS (\%)} & \bf{ROA coverage (\%)} \\ \hline
CLOUDFLARENET   & 28.3 &  98.2                 \\ \hline
AMAZON-02       & 14.4 &  98.9                  \\ \hline
AKAMAI-ASN2     & 7.1  &  100               \\ \hline
NSONE           & 3.1  &  50.0                      \\ \hline
GODADDY-DNS     & 2.8  &  100               \\ \hline
UltraDNS        & 2.8  &  11.1                     \\ \hline
Google, US      & 2.7  &  100               \\ \hline
Total           & 61.2 &   -                        \\
\hline
\end{tabular}
\caption{Top nameserver hosting providers and the proportion of their network prefixes with valid ROA.
}
\label{table:ns-prov-distr}
\end{table}

\bheading{Implications for the attack surface.}
These findings outline several new opportunities for BGP hijacking using DNS. DNS nameservers present a higher number of potential hijack targets with greater geographical diversity, affording attackers more chances to conduct a hijack localized to a CA's vantage point(s) and nameserver that evades global BGP detection. Overall, the additional attack targets may allow some adversaries to succeed in hijacking a domain when they would not have been able to target the domain's A records directly. 
Next, we will use Internet topology simulation and a quantitative metric to evaluate the security of multiVA deployments under the enlarged surface.  

\section{Internet Topology Simulations: A Multi-Faceted Approach}\label{sec:sim}

Given the potential target IP addresses for each domain from our DNS measurements, we estimate each domain's vulnerability to BGP hijacks. We divide this task into two parts: 1) Internet-wide topology simulations that model the routing behavior of every IP prefix on the Internet under various RPKI conditions and simulated attacks from 1\,K randomly chosen adversaries and 2) domain resilience computation using data about target IPs for each domain and the routing information computed by the simulator. Overall, our approach allows us to understand how vulnerable a domain is to BGP attacks.

\subsection{Simulation Methodology}
\label{sec:simulation_methodology}

To measure the impact of BGP attacks on the PKI, we run global equally-specific BGP simulations. We model Gao-Rexford route preferences~\cite{gao2001gao_rexford} over the CAIDA AS topology model~\cite{caida}, and perform IP prefix-level (instead of AS-level) simulations using RIB data from RouteViews~\cite{routeviewUO} and RIPE RIS~\cite{ripe_ncc_ris}. Given a specific adversary AS and a specific victim AS, we simulate whether a potential set of vantage points would route data to the victim or the adversary. 

\bheading{Novel collection of cross-cloud datacenter peering connections.}
In addition to using prefix-level simulations, we needed to accurately model the BGP connections of different potential vantage points used in our simulation. Because public BGP data sources only have a partial view of global routing, many (particularly peering) links that are heavily used by cloud data centers are missed in both public topologies and BGP data. We augment these data sources with neighbors found by running traceroute and the bdrmap~\cite{bdrmap} tool at all cloud vantage points considered. This work is the first to consider multiVA deployments spread across multiple cloud providers (as previous work considered vantage points hosted solely in AWS datacenters~\cite{birgelee2021experiences}). We collected bdrmap data from 19 distinct datacenters spread across three cloud providers (AWS, GCP, and Azure) and four continents (Europe, Asia, North America, and South America).\footnote{Let's Encrypt's two primary data centers in Denver and Salt Lake City operate out of a provider that does not lease cloud services preventing us from running bdrmap. We use public BGP data for these providers.} 
By collecting bdrmap data across different cloud datacenters, we can determine the optimal providers for hosting vantage points and consider the effectiveness of cross-cloud multiVA deployments.

\bheading{Simulation of RPKI-enabled ROV.} 
In addition to non-RPKI simulations, we also ran RPKI-enabled simulations for every prefix in the route table. In the RPKI simulations we modeled the adversary's announcement as having the true victim's origin ASN prepended to evade RPKI-based Route Origin Validation (or ROV). All of Let's Encrypt's cloud vantage points perform ROV \cite{awsrov, gcprov, azurerov}. Any announcements for an IP prefix covered by RPKI which does not contain the true origin ASN will be filtered and not used by Let's Encrypt's vantage points. In this event (where no remote vantage points use the adversary's route) the adversary cannot succeed in obtaining a certificate. 
Thus, \textbf{the only way for an adversary to succeed in an equally-specific prefix attack against an RPKI-protected prefix is to evade ROV and prepend the true origin ASN to its announcement}. While this is a viable strategy (which we have seen in the wild to evade ROV~\cite{coinbase2022celer_bridge}), it does cause the adversary's malicious BGP announcement to be longer and less-preferred which we model in our RPKI-enabled simulations.

Our resilience processor is also capable of incorporating the concrete usage of ROAs and ROVs in the Internet today. It does this by loading both the RPKI and non-RPKI simulations and then determining on a per-IP basis which set of simulation results to use based on one of three different modes:

\textbf{1. No RPKI adoption:} This mode assumes that no IP addresses are covered by RPKI and the non-RPKI simulation results are used. We include it as a baseline and for comparison with previous work (which did not consider RPKI).
    
 \textbf{2. Current RPKI adoption:} This mode uses data from Routinator~\cite{routinator} to determine which IP addresses are currently covered by RPKI, enabling the resilience processor to use the appropriate (i.e., RPKI or non-RPKI) simulation results for each IP address. Routinator is open-source RPKI software (designed to help routers perform ROV) that downloads RPKI data from the five RIR trust anchors and produces a dataset of IP prefixes covered by RPKI, identical to the procedure used by real routers when producing filtering rules for ROV. 
    
 \textbf{3. Full RPKI adoption:} This mode represents the "best case" scenario assuming RPKI adoption has expanded to the entire routing table. In this mode only the RPKI simulation results are used. While current RPKI adoption models the current routing table, we expect that RPKI adoption will increase over time, which will improve the resilience of domains and bridge the gap to  resilience of the full RPKI adoption results. 

Finally, we simulated BGP attacks on every prefix seen in the public BGP data from a set of one thousand adversaries randomly sampled from all ASes in the route table.

\subsection{Defining a Domain's Resilience}
\label{sec:resilience}

Considering the output of the Internet topology simulations and our DNS lookups, we computed an effective resilience for each domain contained in the DNS data. The effective resilience measures the fraction of ASes on the Internet that are topologically \emph{incapable} of acquiring a malicious certificate for a given domain by launching an equally-specific BGP attack~\cite{birgelee2021experiences, lad2007resilience, wubbeling2016resilience, brandt2021dns_resilience}. Intuitively, domains that have weak BGP connectivity have a lower effective resilience as it is highly likely an adversary will succeed in its attack. Domains with stronger BGP connectivity have a high effective resilience. In addition to domain connectivity, the number of CA vantage points and quorum policy impact the effective resilience of domains as stricter quorum policies and more vantage points allow a CA to detect more BGP attacks and reduce the likelihood that an attack will be viable.

We compute the effective resilience of each domain by considering each vantage point we studied and seeing which adversaries are able to successfully hijack communication from that vantage point to \emph{any} target IP address in $Q^+(d)$ (defined in \secref{sec:dns-define}) for the domain in question (see Appendix Algorithm~\ref{alg:resilience}). If, based on our simulations, an adversary succeeds in hijacking traffic from a vantage point to either an A record IP address or a vulnerable DNS nameserver IP address, we consider the adversary to be capable of hijacking the validation request from that vantage point. We then consider the quorum policy and compute, for a given domain, which adversaries are capable of hijacking validation from enough vantage points to satisfy the quorum policy. If an adversary cannot hijack enough validation requests to satisfy the quorum policy, the domain is considered to be resilient to attacks from that adversary. The fraction of randomly sampled adversary ASes that a domain is resilient to is the domain's resilience value. We present the mathematical definition of resilience in Appendix~\ref{sec:app:resilience}.
\section{Quantifying the Impact of Real-World Dynamics on multiVA}
\label{sec:results}

We run resilience simulations on the DNS dataset in \secref{sec:le-domain-profile} to quantify the change in resilience resulting from considering DNS nameservers, RPKI-ROV, and DNSSEC in the BGP hijack attack surface. Our simulations aimed to model a wide array of possible CA vantage point configurations as well as Let's Encrypt's current multiVA deployment.
Our results show that DNS significantly broadens the attack surface available to a BGP hijacking adversary: for a single vantage point, DNS translated to an average resilience drop of over $26\%$. At the same time, RPKI-ROV records provided a resilience gain of $9.9\%$ on average across all configurations tested.
 
\subsection{Re-evaluating Let's Encrypt Present MultiVA Deployment}
\label{sec:le-dns-analysis}

\begin{figure*}
    \centering
    \begin{subfigure}{0.35\textwidth}
        \includegraphics[width=\textwidth]{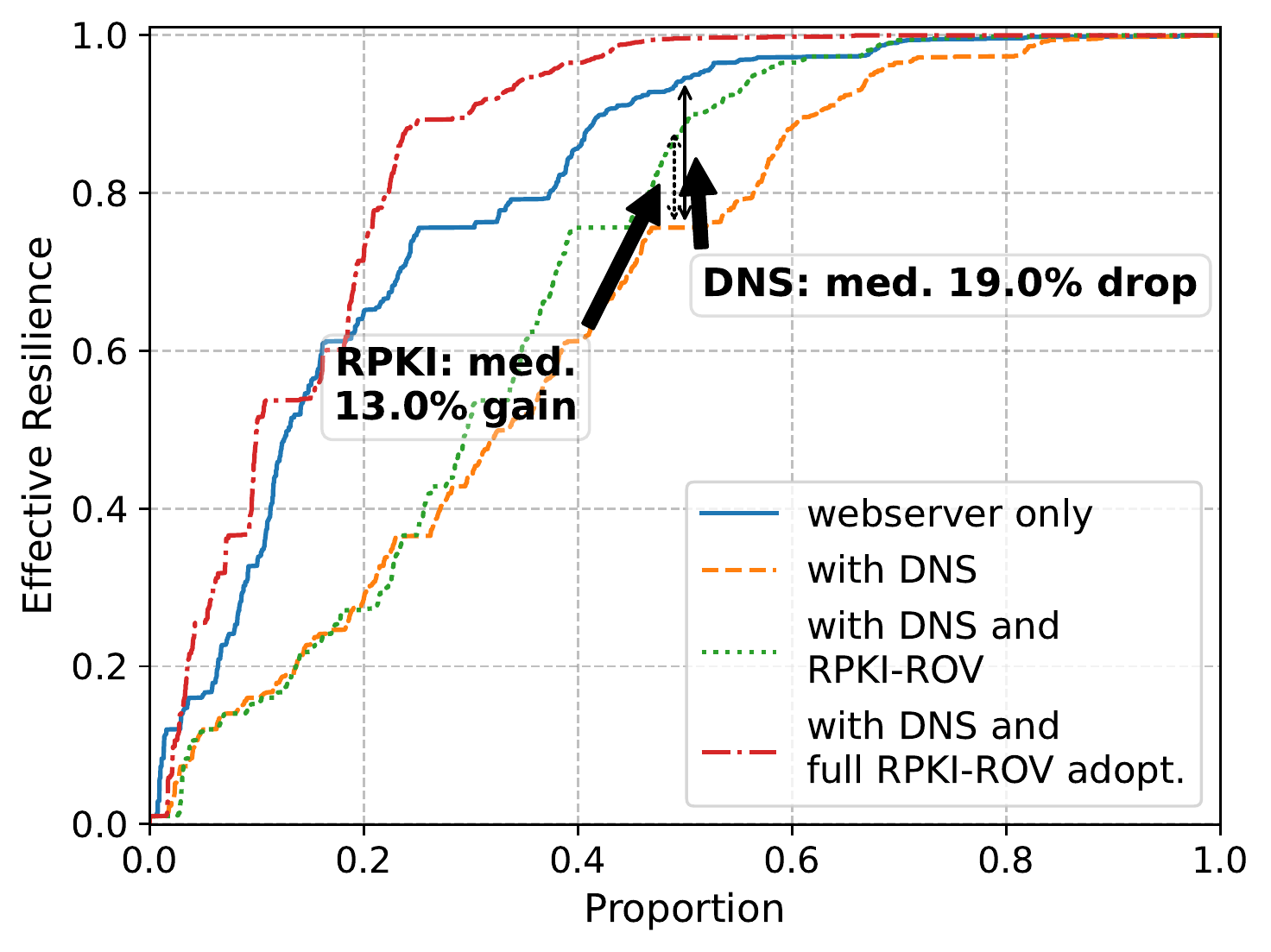}
        \caption{multiVA}
        \label{fig:res-le-cdf}
    \end{subfigure}
    \begin{subfigure}{0.35\textwidth}
        \includegraphics[width=\textwidth]{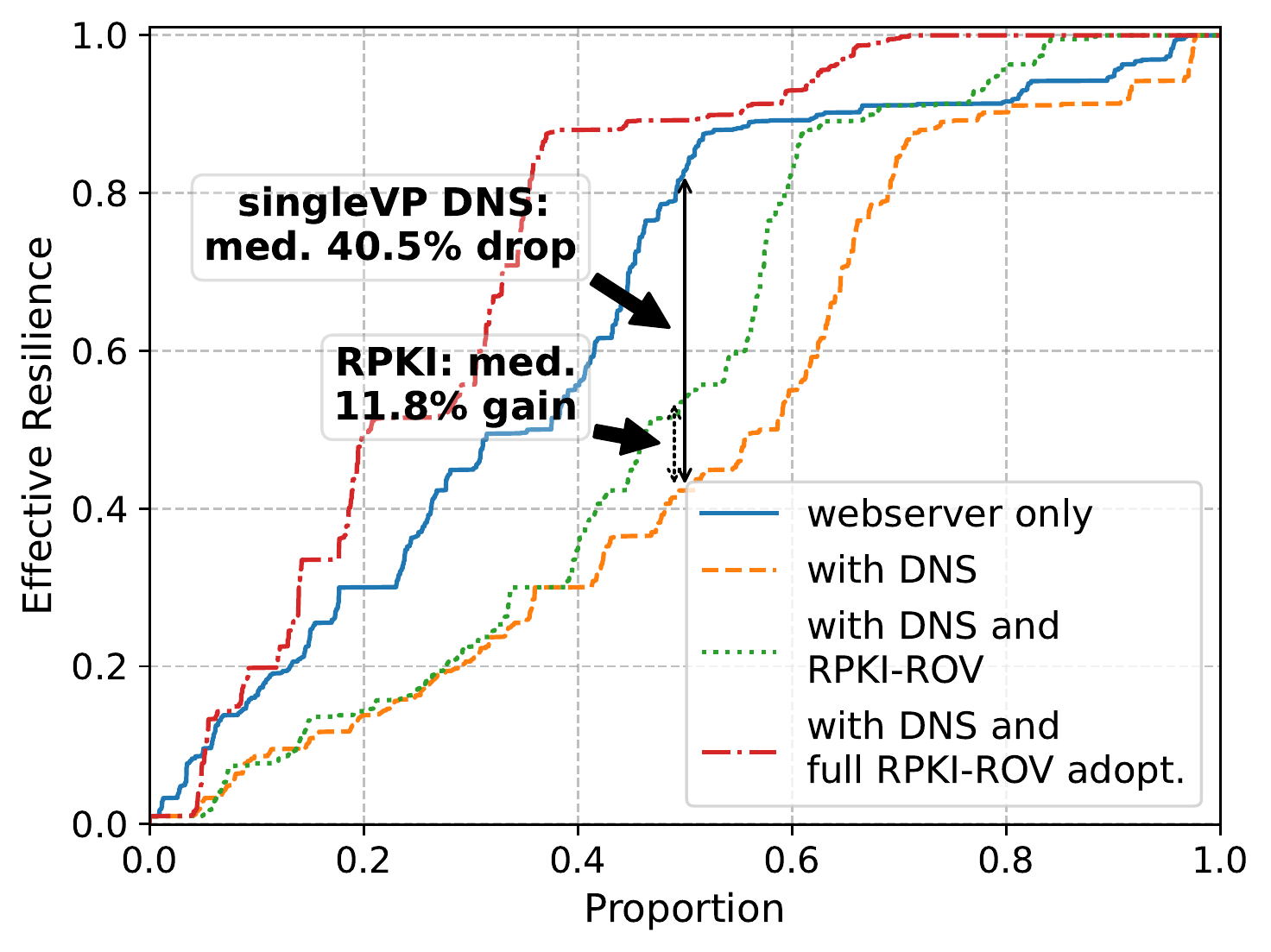}
         \caption{singleVA}
        \label{fig:res-singlevp-cdf}
    \end{subfigure}
    \caption{Comparing the resilience of \textbf{(a)} Let's Encrypt's multiVA  and \textbf{(b)} singleVA deployment under different conditions. }
    \label{fig:le-deployment}
\end{figure*}

\bheading{Resilience of LE deployment drops when considering DNS.}
\figref{fig:res-le-cdf} compares the resilience of Let's Encrypt current deployment of multiVA domain validation when considering case (1) BGP hijacks on only the domain's webserver; case (2) BGP hijacks on the domain's webserver and non-DNSSEC signed-nameservers; case (3) BGP hijacks on all targets in (2) while accounting for the usage of RPKI-ROV by the ASes hosting Let's Encrypt's vantage points. 
Including DNS as an hijack target reduces the median domain resilience by over $20\%$, from $94.6\%$ to $75.6\%$. Approximately one-third of domains sampled had a resilience of less than 0.5, meaning that they are can be hijacked by the adversary in the average case. The result suggests DNS indeed enlarges the attack surface of multiVA and affects its security.  

We also plot the effect of case (1) and case (2) for Let's Encrypt's original single vantage point-deployment prior to 2020 in \figref{fig:res-singlevp-cdf}. The drop in resilience due to NS records is dramatically higher for the single vantage point case: here resilience drops by over 33\%. Thus, although multiVA is not a complete remedy for DNS-based BGP hijacks, it minimizes the resilience drop compared to the singleVA case.

\label{sec:le-rpki-analysis}

\bheading{RPKI improves resilience even more so for multiVA.}
The resilience degradation presented in \secref{sec:le-dns-analysis} is pessimistic and does not consider the enhancement in routing security attributable to RPKI-ROV deployment.  
When considering the deployed ROA-ROV on top of the DNS model, the resilience increases significantly, from $75.6\%$  to $88.6\%$. When considering the situation of full RPKI-ROV adoption (i.e., where all prefixes in the routing table have a valid ROA registered), the median domain resilience shoots up to 99.6\%. It is interesting to note that the resilience benefit from RPKI-ROV is significantly higher for multiVA than the single vantage point case: the median domain resilience under the full RPKI-ROV adoption regime for single VA is only 89.2\%.
The variance in resilience measurements highlights the importance of considering both DNS and RPKI data in the domain resilience estimation. Still, the registration of RPKI records for domain-hosting prefixes is largely out of the CA's control; in light of this, we explore more multiVA configurations in \secref{sec:le-res-improve} to augment CA resilience.

\bheading{Takeaways.}
Our results suggest it is necessary for future research on multiVA security to take deployed security measures into account in evaluation. This will produce a more realistic and accurate estimation of the security of a multiVA deployment. To better protect their domains, domain owners should choose hosting providers that enable RPKI~\cite{isbgpsafeyet} and turn on DNSSEC if possible. 

\subsection{Strategies to Improve MultiVA Security}
\label{sec:le-res-improve}

In this section, we explore various techniques to improve the security of multiVA deployments. We consider four factors that a multiVA deployment can control: the number of vantage points, vantage point geographic location, vantage point cloud provider, and CA quorum policy to issue certificates. We compute the resilience of Let's Encrypt multiVA under 11,110 combinations of these parameters, and summarize our observations and takeaways from the optimal configurations. We use the resilience of the current Let's Encrypt deployment (while modeling DNS and RPKI) as the baseline.

\begin{table}[t]
\centering
\footnotesize
\begin{tabular}{|r|r|r|r|}
\hline
\multicolumn{1}{|c|}{\textbf{\begin{tabular}[c]{@{}c@{}}\# Additonal \\ VPs\end{tabular}}} & \multicolumn{1}{c|}{\textbf{Quorum}} & \multicolumn{2}{c|}{\textbf{Resilience (median)}} \\ \cline{3-4}

& & \multicolumn{1}{c|}{\textbf{Single-cloud}} &\multicolumn{1}{c|}{\textbf{Multi-cloud}} \\ \hline

\textbf{0} & $\mathbf{n-1}$ & \textbf{88.6} & ---            \\ \hline
1          &  n-1           &   93.2        & 93.2           \\ \hline
2          &  n-1           &   94.6        & 97.5    \\ \hline
3          &  n-1           &   95.8        & 97.7    \\ \hline
4          &  n-1           &   96.9        & 97.8            \\ \hline
5          &  n-1           &   97.0        & 98.6            \\ \hline \hline
0          &  n           &   96.5        & ---            \\ \hline
1          &  n           &   97.5        & 98.4   \\ \hline
2          &  n           &   98.2        & 99.3 \\ \hline
\end{tabular}
\caption{Effect on Resilience with varying vantage point counts/locations and quorum policies, assuming current RPKI-ROV deployment. (Boldface indicates LE's present deployment.) Extra VPs are \textit{in addition} to LE's current deployment of 3 remote VPs ($n = 3$).
}
\label{tab:strategy}
\end{table}

\begin{figure*}
    \centering
    \begin{subfigure}{0.306\textwidth}
        \includegraphics[width=\textwidth]{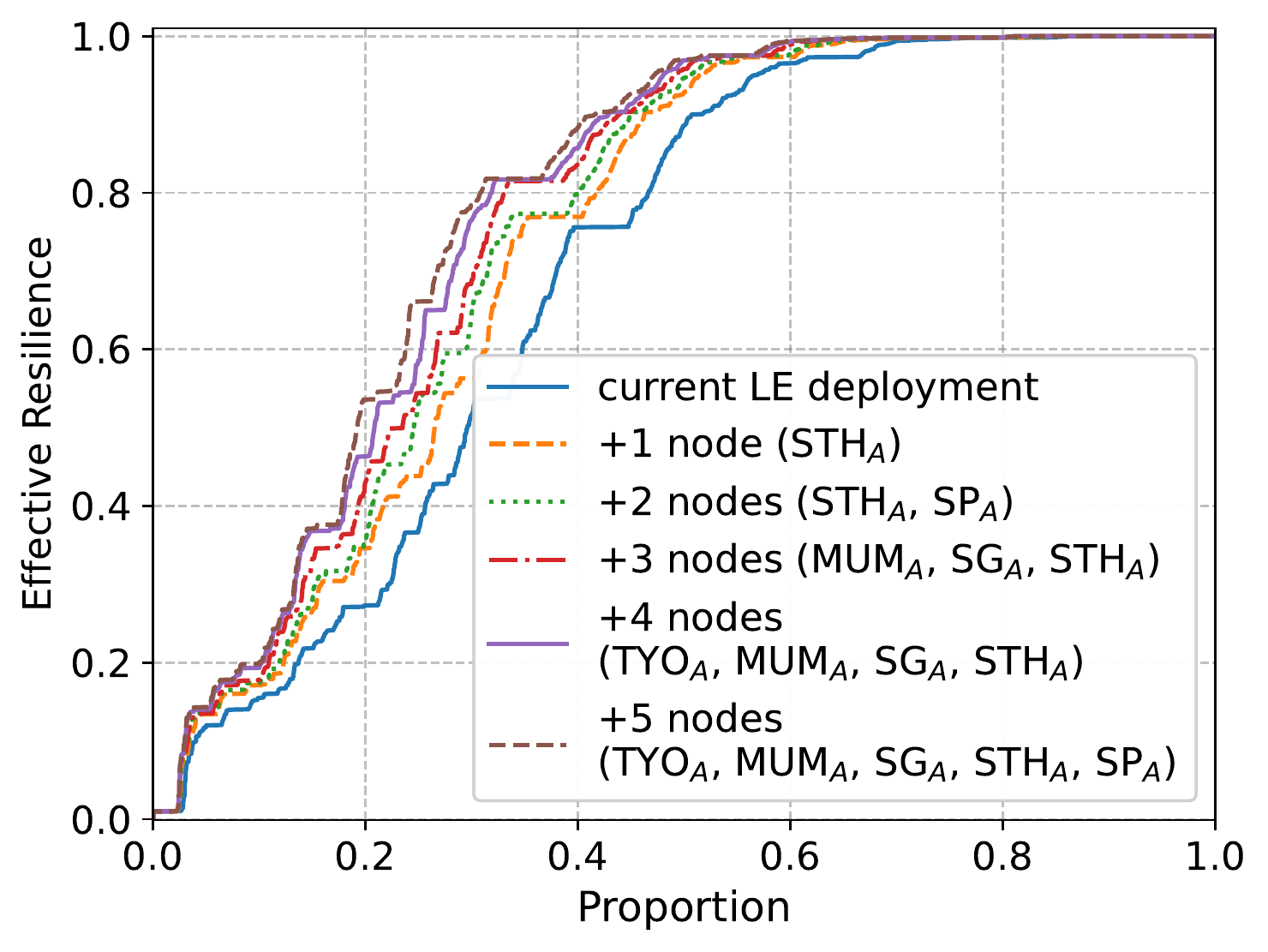}
        \caption{}
        \label{fig:le-deploy-aws}
    \end{subfigure}
    \hfill
    \begin{subfigure}{0.306\textwidth}
        \includegraphics[width=\textwidth]{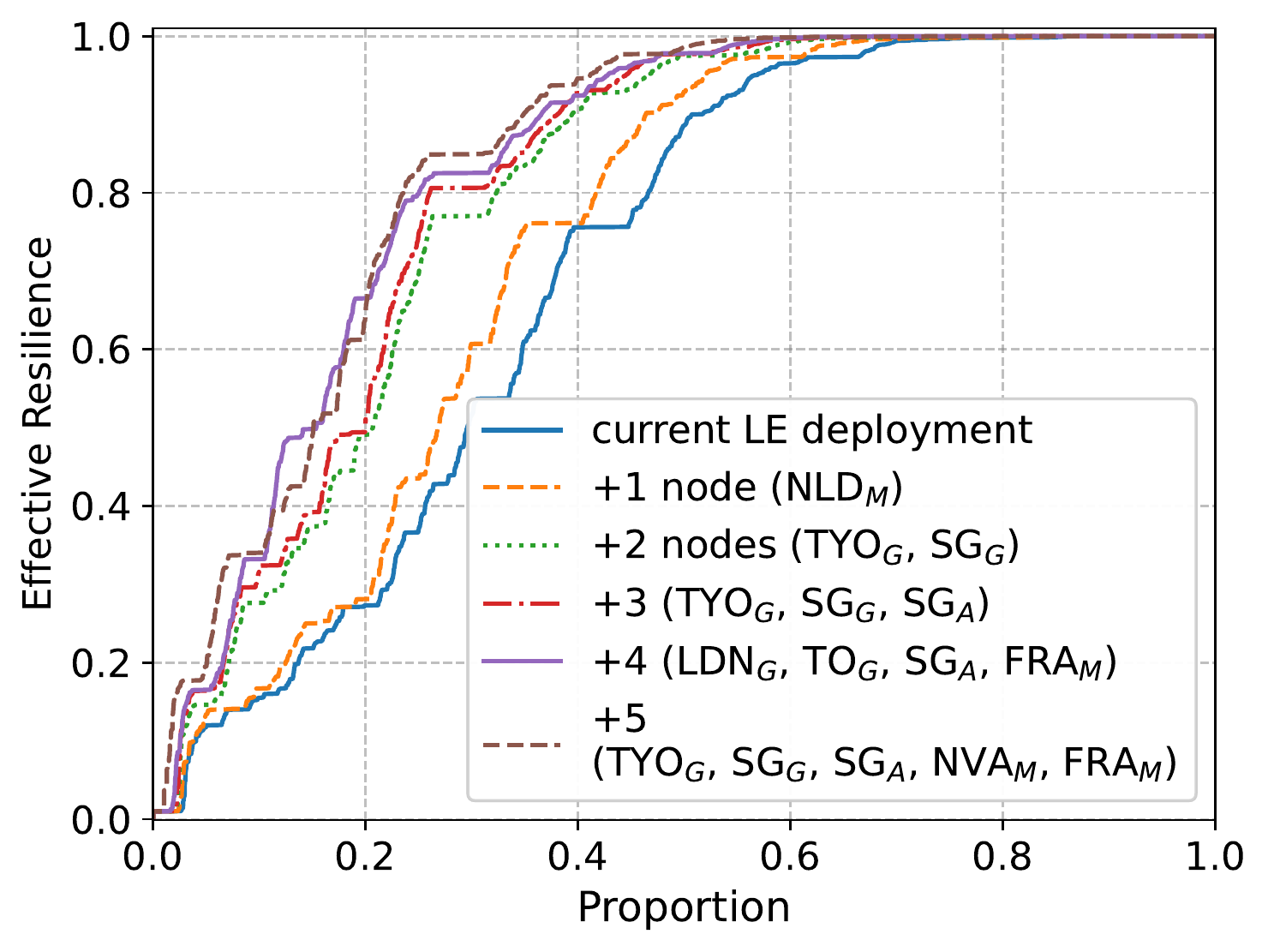}
        \caption{}
        \label{fig:cross-cloud-res-cdf}
    \end{subfigure}
    \hfill
    \begin{subfigure}{0.368\textwidth}
        \includegraphics[width=\textwidth]{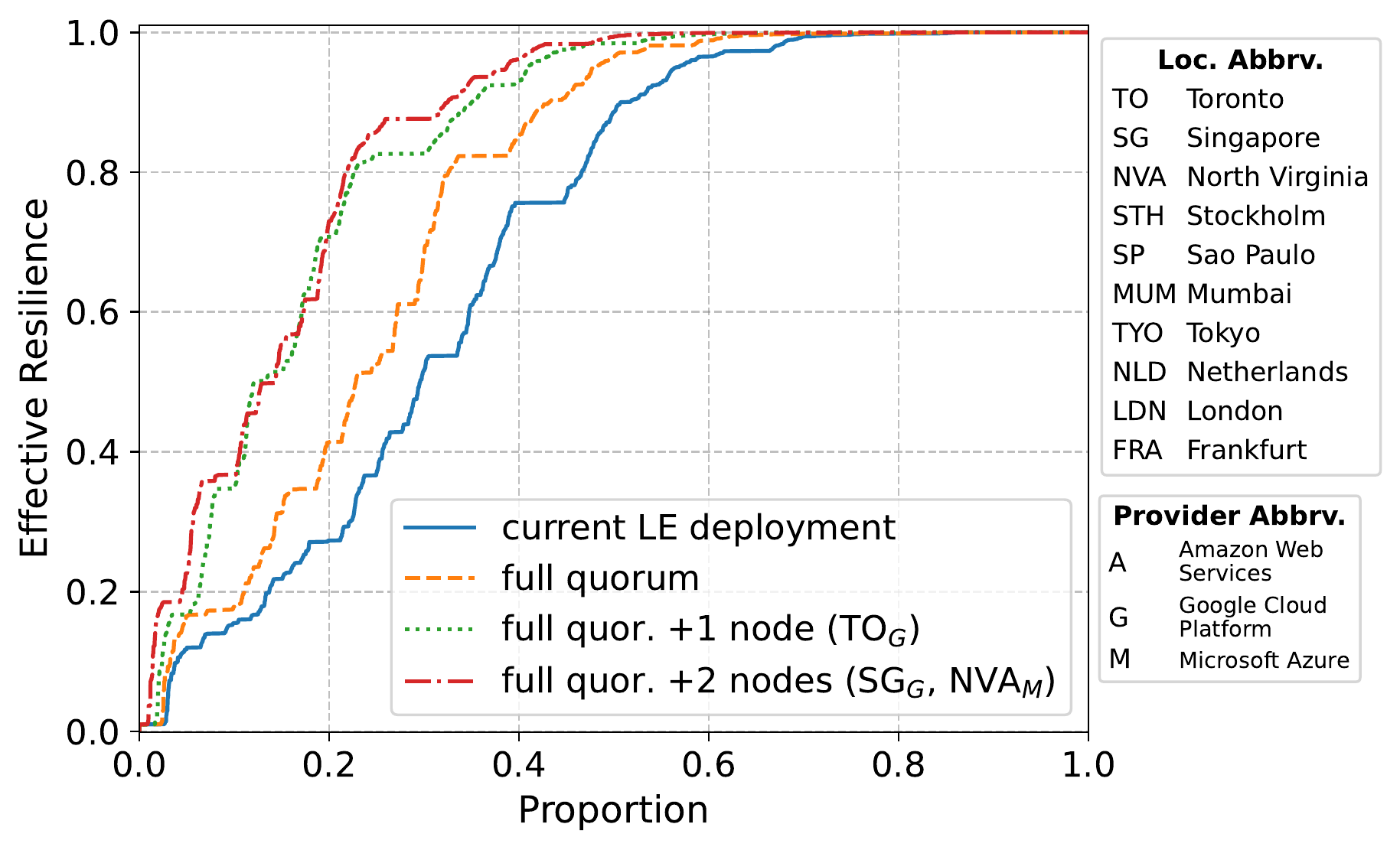}
        \caption{}
        \label{fig:full-quorum-res}
    \end{subfigure}
    \caption{Comparing resilience of Let's Encrypt's deployment when \textbf{(a)} adding AWS vantage points; \textbf{(b)} adding cross-cloud vantage points; \textbf{(c)} changing to a full quorum policy.}
    \label{fig:triple-resilience}
\end{figure*}

\bheading{Adding vantage points using AWS.}
Let's Encrypt's current deployment of multiVA has remote vantage points hosted exclusively in AWS datacenters (in addition to its primary data center hosted by the datacenter provider Flexential). Let's Encrypt chose this approach to avoid the complexity involved with deploying vantage points in multiple cloud providers, which entails significant engineering overhead \cite{birgelee2021experiences}. Thus, a logical and low-cost extension to LE's deployment would be to simply add additional AWS vantage points. We find that a single cloud provider deployment requires a significant number of vantage points to be added for increased effectiveness (see \figref{fig:le-deploy-aws}). For example, the AWS deployment flatlines at a maximal $97\%$ resilience after adding 5 additional vantage points to Let's Encrypt's deployment.

Considering all the AWS vantage point surveyed, we found the optimal AWS vantage points (i.e., those with the largest resilience increase) to be concentrated in Asia (Tokyo, Mumbai, and Singapore), along with another perspective in Northern Europe (Stockholm) and South America (Sao Paulo).

Adding the first additional AWS vantage point (with the same quorum policy) shows some promise, improving resilience from $88.6\%$ to $93.2\%$. We posit this is because another vantage point in Europe is highly effective given Let's Encrypt's current deployment and quorum policy. Let's Encrypt's Frankfurt vantage point provides greater routing diversity than its two US vantage points~\cite{birgelee2021experiences}. However, Let's Encrypt's quorum policy allows a vantage point to fail, meaning that attacks detected solely by the Frankfurt vantage point still succeed. Adding Stockholm as a vantage point supports Frankfurt in representing the European vote and can help protect against attacks where the European perspective is crucial.

On the other hand, adding another AWS vantage point only improves resilience to $95.1\%$ and a third only brings resilience to $96.6\%$. A fourth vantage point yields an incremental resilience gain of $0.3\%$ more to $96.9\%$, and a fifth maximizes at $97.0\%$. Although the resilience benefit of adding more EC2 nodes saturates, this approach still does increase resilience by nearly 10\% while requiring minimal engineering setup and management costs. 

\bheading{Adding vantage points using additional cloud providers.}
A more substantial resilience gain can be achieved by adding additional vantage points in alternate cloud providers. If one more vantage point is added, Microsoft Azure offers the optimal next vantage point location in West Europe (Amsterdam), which improves resilience by a slight 0.1\% over the first optimal EC2 location.  If two additional vantage points are deployed, the optimal configuration becomes GCP, with nodes in Asia Northeast (Tokyo) and Asia Southeast (Singapore); this bumps up median resilience to 97.5\% representing a 8.9\% improvement over Let's Encrypt's current deployment and a 2.9\% improvement over an all-AWS based deployment with the same number of vantage points. These gains can be further maximized with three additional vantage points spread across AWS and GCP, which scores a 97.7\% resilience figure. Finally, a deployment using 5 nodes across all three cloud providers studied reaches a resilience of 98.6\% (see~\figref{fig:cross-cloud-res-cdf}).

\begin{table}[t]
\centering
 \small
\begin{tabular}{|r|c|c|c|} 
 \hline
 \bf{Provider} & \bf{AWS} & \bf{GCP} & \bf{Azure} \\
 \hline
 AWS & 0.34 & 0.10 & 0.15 \\ \hline
 GCP & 0.10 & 0.54 & 0.25 \\ \hline
 Azure & 0.15 & 0.25 & 0.78 \\ 
  \hline
\end{tabular}
\caption{Average fraction of overlapping peers between datacenter pairs of different providers. Because of homogeneity within providers, pairs of datacenters from different providers have a lower average overlap than pairs from a single provider.}
\label{table:overlapping}
\end{table}

The advantage of a cross-cloud deployment can be found in the peers for various datacenters found with bdrmap. If two datacenters have a higher percentage of overlapping peers, they are likely to have a higher degree of similarity in BGP routing tables as routing by cloud providers relies heavily on peer-learned routes~\cite{arnold2020cloud_provider_connectivity}. We calculated the fraction of overlapping peers between two datacenters (with sets of peers $P_1$ and $P_2$) as $\frac{|P_1 \cap P_2|}{max(|P_1|, |P_2|)}$. We computed this overlap fraction between all pairs of datacenters we studied and grouped the results by the cloud providers. We found that on average datacenters of the same provider had a higher overlapping peer fraction than datacenters of different providers (see \tabref{table:overlapping}). This explains why a cross-cloud deployment can achieve more route diversity even with a smaller vantage point count.

\bheading{Changing the quorum policy.}
Another approach to improve resilience that requires relatively little engineering resources is to change the quorum policy from allowing one vantage point to fail to requiring complete consensus among vantage points. This makes the adversary's task significantly harder because the adversary now has to successfully hijack traffic from all of Let's Encrypt's vantage points because a single vantage point routing to the victim will prevent issuance. While this does significantly improve resilience, it also incurs additional false positives (benign failures in the absence of an attack) as discussed in prior work~\cite{birgelee2021experiences}. The CA needs to strike a balance between security and accuracy. One way to achieve this balance while still strengthening the quorum policy is to encourage users and hosting providers to wait until DNS propagation is complete before requesting a certificate as partial DNS propagation is one of the leading causes of false positives~\cite{birgelee2021experiences}. In addition, distinguishing benign failure modes (like NXDOMAIN errors at remote vantage points that do not imply malicious activity) from the types of responses that can arise during an attack has the potential to reduce the false positive rate and make changing the quorum policy a viable option. 

Simply changing the quorum policy alone (with the current set of vantage points) improves median domain resilience to 96.5\%. This significant 7.9\% increase is likely due in part to the dynamics of Let's Encrypt's current deployment where Frankfurt is the only non-US vantage point and can be ignored in the current quorum policy (allowing quorum to be met with only US-based vantage points). If we further consider changing the quorum policy and adding one additional vantage point, \textbf{a 98.4\% resilience can be achieved (which outweighs the extra attack surface introduced by DNS)}. Further, adding two vantage points under the full quorum policy improves resilience to 99.3\% (see \figref{fig:full-quorum-res}).

\section{Related Work}
\label{sec:related}

Related work falls into three main categories:  network attacks on domain control validation, broader BGP attacks and defenses, and studies of the DNS infrastructure.

\bheading{Network attacks on domain control validation.}
\label{sec:related:domain_validation}
Domain control validation is a critical service and is vulnerable to several different types of network attacks. Gavrichenkov \emph{et al.}  discussed the vulnerability of domain control validation to BGP attacks~\cite{gavrichenkov2015breaking} and Birge-Lee \emph{et al.} were the first to ethically demonstrate such attacks in the wild~\cite{birge2018bamboozling}. In addition, through a collaboration with Let's Encrypt, Birge-Lee \emph{et al.} designed and deployed multiVA to protect domain control validation from these attacks~\cite{birgelee2021experiences}.

Several works (like those by Birge-Lee \emph{et al.}~\cite{birge2018bamboozling,birgelee2021experiences}) have looked at the resilience of domains to BGP attacks. However, these earlier works explored a significantly different (and limited) attack surface and did not consider the impact of DNS or RPKI. A more recent poster on DNS attacks on domain validation did a preliminary investigation into the impact of DNS~\cite{brandt2021dns_resilience} but did not fully account for the DNS attack surface by not considering full DNS lookup graphs, DNSSEC, geographically distributed DNS lookups, or repeated DNS lookups. Furthermore, the work by Brandt \emph{et al.}~\cite{brandt2021dns_resilience} does not consider RPKI, or vantage points beyond Let's Encrypt's current vantage points, and it does not consider any recommendations for how multiple vantage point validation can be improved through expansion to different cloud providers or changes to the quorum policy.

Beyond BGP attacks, domain control validation is vulnerable to other classes of network attacks. DNS attacks (without using BGP) can also be used to attack domain control validation~\cite{dai2021downgrade}. In fact, Brandt \emph{et al.} recommend  a multiple-vantage-point based approach as a mitigation for DNS attacks on domain control validation~\cite{brandt2018dv++}. Dai \emph{et al.} discuss DNS attacks against Let's Encrypt's multiple vantage point validation and demonstrate a novel technique to force all of Let's Encrypt's vantage points to query the same authoritative nameserver~\cite{dai2021downgrade}. We consider the implications of this work in our analysis by assuming that an adversary capable of compromising \textbf{any} non-DNSSEC authoritative nameserver associated with a victim's domain is capable of obtaining a fraudulent certificate.

\bheading{BGP attacks and defenses.}
Our work builds upon important prior works in BGP security. 
The notion of domain resilience (also similar to the constructions in~\cite{birge2018bamboozling, birgelee2021experiences, birgelee2022sbas,sun2015raptor}) was originally adapted from AS-resilience as defined by Lad \emph{et al.} ~\cite{lad2007resilience}. Additionally, the underlying algorithm behind our simulation engine is based on the algorithm presented by Gill \emph{et al.}~\cite{gill2012quicksand}.

In addition to impacting our tooling, several prior works in BGP security have informed our analysis. Crucially we incorporate RPKI~\cite{rpki} as it is currently deployed in the routing table. We also take note of Gilad \emph{et al.}'s discussion of the max-length attribute in RPKI~\cite{RPKI_max_length} and how its misuse can undermine the security of RPKI. However, the vulnerability introduced by improperly configured max-length attributes in ROAs is fundamentally a sub-prefix attack,
which is outside our threat model. Moreover, the vast majority of prefixes covered by RPKI are not impacted by this type of misconfiguration and this attack vector can be mitigated with proper RPKI configuration~\cite{gilad2017rpki_deployment_security}. We further acknowledge the impact of BGP security practices like prefix filtering (as recommended by the MANRS project~\cite{MANRS}) and peer locking~\cite{snijders2016peerlock}. While both these practices have a positive impact on network security, we do not explicitly consider these in our work. The prefix filtering techniques outlined by MANRS are not fool-proof and have been evaded by real-world attacks~\cite{coinbase2022celer_bridge} and peer locking is only viable for ASNs directly connected a handful of participating providers and not the topology at large~\cite{snijders2016peerlock}.

\bheading{DNS Mapping.}
 Prior work~\cite{ram2005transtrust} introduced the notion of full-graph DNS and the implications of transitive trust between nameserver delegations for hijacking DNS names. Other work~\cite{pappas2004impact, deccio2009quality, deccio2010measuring} has leveraged full-graph DNS resolution to understand DNS resolution dependency to study the security, availability, and robustness of DNS. Unfortunately, the full-graph resolvers in prior work do not record DNSSEC, consider the names in glued records as dependencies, and none of them are open-sourced. We therefore develop our own resolver that provides multiple new features, such as geographically-distributed lookups and collection of DNSSEC information. 
\section{Conclusion}
\label{sec:background}

We develop a novel analysis framework capable of considering the most complete attack surface of BGP hijacks on CAs to date. 
Our work provides rigorous quantitative data and analysis about the effectiveness of multiVA under real-world conditions. Our promising results,  serving as critical empirical evidence, speak strongly for the adoption of multiVA by the CA and Browser Forum. Furthermore, our analysis of different multiVA configurations can help optimize the deployment of current and future multiVA deployments. 
Moreover, our analysis framework that accounts for the attack surface introduced by DNS, as well as protections offered by RPKI, has broad applicability to other network security and privacy domains beyond the PKI. We present our analysis framework as a primary contribution of this paper. Collecting enough data to accurately model the complexities of routing in the context of certificate issuance (probing DNS nameservers, inferring cloud provider peers, internet-scale topology simulations, etc.) entailed significant financial resources and engineering efforts which can benefit follow-up research in the security community. For example, our analysis methodology can aid developers of anonymity systems and cryptocurrencies to accurately assess security against BGP attacks~\cite{sun2015raptor,apostolaki2017bitcoin}. 

\section*{Acknowledgments}
We would like to thank our collaborators at Let's Encrypt including Josh Aas, Jillian Karner, Aaron Gable and James Renken for providing data access and guidance on this project. We thank Hyojoon Kim, Sophia Yoo, and Mona Wang Li for their feedback on the paper. 

\bibliographystyle{plain}
\bibliography{paper}

\appendix

\begin{figure}[t]
\centering
\includegraphics[width=0.8\linewidth]{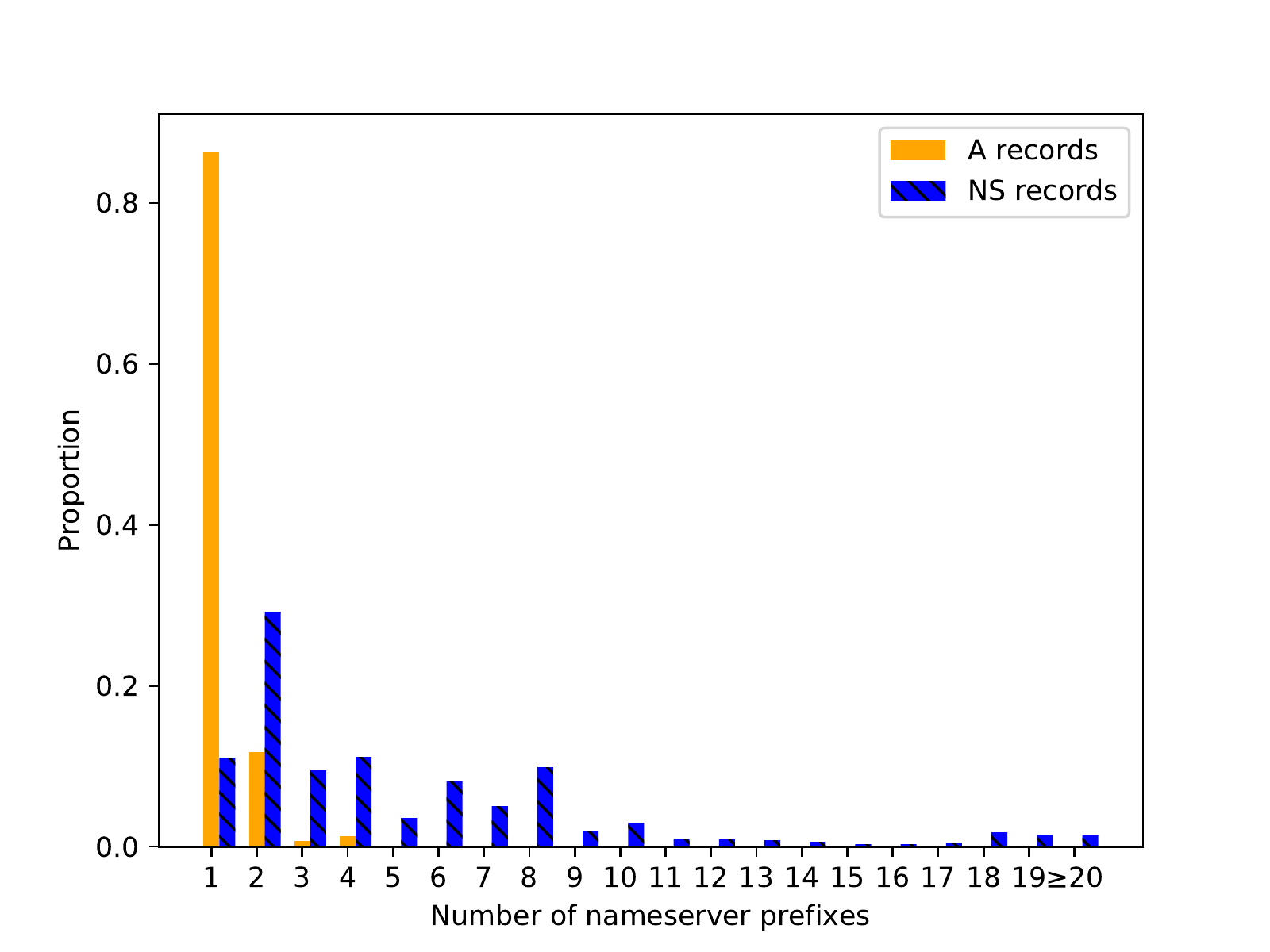}
\caption{
Distribution of number of prefixes for A and NS records, with respective median of 1.0 and 3.0 prefixes. 
}
\label{fig:prfx-distr}
\end{figure}

\begin{algorithm}[t]
    \SetKwInOut{Input}{Input}
    \SetKwInOut{Output}{Output}
    
    \underline{function getResilience $(d)$}\;
    \Input{Domain name $d$}
    \Output{res($d$), where $0 \leq \text{res}(d) \leq 1.0$}
    $successCount \gets 0$\;
    $Adversaries \gets \{\text{Sampled ASes}\}$\;
    \ForEach{ $a \in Adversaries$}{
        $V \gets \{\}$\;
        \ForEach{ $vp \in VantagePoints$}{
             \lIf{$a$ \text{is capable of hijacking traffic from} $vp$ \text{to ANY IP addresses in} $Q(d)$}{
             $V \gets V \cup vp$
             }
            }
            \lIf{quorum(V) == true}{
                $successCount$++
            }
            
        }
    \KwRet{$1 - \frac{successCount}{|Adversaries|}$}\;
    \caption{Resilience computation for a domain $d$. $Q(d)$ are the target IP addresses of domain $d$ as defined in Section~\ref{sec:dns-define}. $V$ is the set of vantage points the adversary can attack.}
    \label{alg:resilience}
\end{algorithm}

\begin{table}[t]
    \centering
    \small
    \begin{tabular}{|c|c|c|}
    \hline
     Provider  &  Location Name & Geographic Location \\
     \hline
     GCP  & asia-northeast1         & Asia Pacific (Tokyo) \\
                            & asia-southeast1         & Asia Pacific (Singapore) \\
                            & europe-west2            & Europe (London) \\
                            & us-east4                & Americas (N.Virginia) \\
                            & us-west1                & Americas (Oregon) \\ 
                            & northamerica-northeast2 & Americas (Toronto)  \\ \hline
     AWS    & eu-west-3               & Europe (Paris) \\
                            & eu-north-1              & Europe (Stockholm)\\
                            & eu-central-1            & Europe (Frankfurt)\\
                            & ap-south-1              & Asia Pacific (Mumbai) \\
                            & us-east-2               & US East (Ohio) \\
                            & us-west-2               & US West (Oregon)\\
                            & sa-east-1               & South America (São Paulo)\\
                            & ap-southeast-1          & Asia Pacific (Singapore)	\\
                            & ap-northeast-1          & Asia Pacific (Tokyo) \\ \hline
                            
     Azure        & germany-west-central    & Frankfurt \\
                            & japan-east-tokyo        & Tokyo, Saitama	\\
                            & us-east-2               & Virginia \\
                            & west-europe             & Netherlands\\
    
    \hline
    \end{tabular}
    \caption{Full list of cloud datacenter locations surveyed with provider-specified location.}
    \label{tab:cloud-prov-location}
\end{table}

\section{Domain Resilience Including Vulnerable DNS IP addresses and A Records}
\label{sec:app:resilience}

We begin by considering our topology simulation results to produce an attack bit (that we will notate as $\alpha$) for each adversary, destination, and vantage point tuple. For a specific adversary ($a$), destination IP prefix under attack ($p$), and vantage point ($v$), $\alpha$ indicates whether the adversary's attack is successful in attracting traffic from the vantage point to the destination (which is indicated by $\alpha(a, p, v) = 1$).

\textbf{Including DNS lookup data.} As prior work has established~\cite{dai2021downgrade}, DNS server selection techniques in popular DNS software (like that used by Let's Encrypt) can be exploited to allow an adversary to \emph{select} which DNS server (or servers in a multi-level DNS lookup) a CA uses. We model this by assuming that an adversary's ability to compromise \emph{any} DNS server in a domain's DNS lookup graph can be used to compromise the lookup of that domain. With this in mind, we analyze our DNS data to extract a target IP list for each domain, vantage point pair that contains 1) the IPs of all non-DNSSEC protected DNS servers in the DNS lookup for the domain from the region of the vantage point, appended with 2) any A records found for the domain in that region.

Given this list of target IPs for each domain, we define $\alpha^*$ as a function of the adversary $a$, the domain name $d$, and the vintage point $v$ as 

$$\alpha^*(a,d,v)=\bigvee_{p \in {IP}(d)}\alpha(a, p, v)$$

where $IP(d)$ is the list of target IP addresses for a name $d$. Under this definition, a domain name is vulnerable to attack at a vantage point (i.e., $\alpha^*(a,d,v)=1$) if \emph{any} of the target IPs for that name at that vantage point are susceptible to hijack by the adversary. 

Given this definition, we incorporate the CAs quorum policy in a similar manner to Birge-Lee \emph{et al.}~\cite{birgelee2021experiences}. By defining a quorum policy $q(\mathcal{W})$ that evaluates to 1 when $\mathcal{W}$ (the set of vantage points controlled by the adversary) is sufficient to issue a certificate, we can define $\alpha^+(a, d, q)$ for an adversary $a$, name $d$, and quorum policy $q$ as:

$$
\alpha^+(a, d, q) = q(\{v \in \mathcal{V} \, | \, \alpha^*(a, d, v) = 1\})
$$

where $\mathcal{V}$ are the set of all vantage points operated by a CA.

Finally, we define the effective resilience for a domain name $d$, a quorum policy $q$, a set of vantage points $\mathcal{V}$ and a set of adversaries $\mathcal{A}$ as:

$$
    \gamma(d, q,  \mathcal{V}, \mathcal{A}) = 1 -  \frac{{\sum\nolimits_{a \in \mathcal{A}}  \alpha^+(a, d, q)}}{|\mathcal{A}|}
$$

Importantly, because we have a liberal definition of $\alpha^+(a, d, q)$ (where $\alpha^*(a, d, v) = 1$ if the adversary capable of attacking communication from vantage $v$ to \emph{any} target IP address of $d$), this formula underestimates resilience and incorporates all nameserver selection techniques discussed in previous work~\cite{dai2021downgrade}.

\section{Cloud Providers and Locations Used}

\tabref{tab:cloud-prov-location} contains the full list of cloud providers and locations modeled in the DNS data collection in \secref{sec:dns} and the simulations in \secref{sec:results}.

\end{document}